\documentclass{aa}  

\usepackage{longtable}
\usepackage{rotating}
\usepackage{graphicx}
\usepackage{txfonts}
\usepackage{tablefootnote}
\usepackage{hyperref}

\usepackage{xcolor}
\definecolor{dark-red}{rgb}{0.9,0.0,0.0}
\definecolor{dark-blue}{rgb}{0.15,0.15,0.9}
\definecolor{dark-green}{rgb}{0.15,0.8,0.15}
\definecolor{medium-blue}{rgb}{0,0,0.9}
\hypersetup{
    colorlinks, linkcolor=red,
    citecolor={dark-blue} , urlcolor={medium-blue}
}
\usepackage{changepage} 

\newcommand{\msini}{m~sin$i$}
\newcommand{\emcee}{\texttt{EMCEE}}
\newcommand{\emp}{\texttt{EMPEROR}}
\newcommand{\teff}{$T_{\rm eff}$\,}
\newcommand{\mj}{$M_{\rm J}$\,}
\newcommand{\msun}{$M_\odot$\,}
\newcommand{\lsun}{$L_\odot$\,}

\newcommand{\rsun}{$R_\sun$\,}

\newcommand{\rstar}{$R_\star$\,}
\newcommand{\logg}{log\,$g$\,}
\newcommand{\feh}{{\rm [Fe/H]}\,}
\newcommand{\vsini}{\textit{v}\,sin\,$i$}
\newcommand{\ms}{m\,s$^{-1}$\,}
\newcommand{\cmsq}{cm\,s$^{-2}$\,}
\usepackage{amstext}

\begin{document}

   \title{Sub-Jupiter Gas Giants Orbiting Giant Stars Uncovered using a Bayesian Framework\thanks{ Based on observations collected at La Silla - Paranal Observatory under programs ID's 085.C-0557, 087.C.0476, 089.C-0524, 090.C-0345, 097.A-9014, 0107.A-9004, 112.2666.001 and through the Chilean Telescope Time under programs ID's CN-12A-073, CN-12B-047, CN-13A-111, CN-2013B-51, CN-2014A-52, CN-15A-48, CN-15B-25 and CN-16A-13.}}

   \author{ J.~S. Jenkins\inst{1,2}
    \and M.~I. Jones\inst{3}
    \and J.~I. Vines\inst{4}
    \and R.~I. Rubenstein\inst{1,5}
    \and P.~A. Pe{\~n}a Rojas\inst{1,2}
    \and R. Wittenmyer \inst{6}
    \and R. Brahm \inst{7,8,9}
    \and M. Tala Pinto \inst{7}
    \and J. Carson \inst{8}
   }

   \institute{Instituto de Estudios Astrof\'isicos, Facultad de Ingenier\'ia y Ciencias, Universidad Diego Portales, Av. Ej\'ercito 441, Santiago, Chile\\\email{james.jenkins@mail.udp.cl}
   \and Centro de Astrof\'isica y Tecnolog\'ias Afines (CATA), Casilla 36-D, Santiago, Chile
   \and European Southern Observatory, Alonso de C\'ordova 3107, Vitacura, Casilla 19001, Santiago, Chile  
   \and Instituto de Astronom\'ia, Universidad Cat\'olica del Norte, Angamos 0610, 1270709, Antofagasta, Chile
   \and United States Fulbright Fellow; Chile Fulbright Commission
    \and  University of Southern Queensland, Centre for Astrophysics, Toowoomba, QLD 4350, Australia
    \and Facultad de Ingenier\'ia y Ciencias, Universidad Adolfo Ib\'a\~nez, Av.\ Diagonal las Torres 2640, Pe\~nalol\'en, Santiago, Chile
   \and Millennium Institute for Astrophysics, Chile
   \and Data Observatory Foundation, Santiago, Chile
   \and Department of Physics \& Astronomy, College of Charleston, 66 George Street Charleston, SC 29424, USA
   }

   \date{Accepted 29/08/2025}

 
  \abstract{Giant stars have been shown to be rich hunting grounds for those aiming to detect giant planets orbiting beyond $\sim$0.5~AU.  Here we present two planetary systems around bright giant stars, found by combining the radial-velocity (RV) measurements from the EXPRESS and PPPS projects, and using a Bayesian framework.  HIP18606 is a naked-eye ($V=5.8~$mags) K0III star and is found to host a planet with an orbital period of $\sim$675~days, a minimum mass (\msini) of 0.8~\mj, and a circular orbit.  HIP111909 is a bright ($V=7.4~$mags) K1III star, and hosts two giant planets on circular orbits with minimum masses of \msini=1.2~\mj and \msini=0.8~\mj, and orbital periods of $\sim$490~d and $\sim$890~d, for planets b and c respectively, strikingly close to the 5:3 orbital period ratio.  Analysis of 11 known giant star planetary systems arrive at broadly similar parameters to those published, whilst adding a further two new worlds orbiting these stars.  With these new discoveries, we have found a total of 13 planetary systems (including three multiple systems) within the 37 giant stars that comprise the EXPRESS and PPPS common sample.  Periodogram analyses of stellar activity indicators present possible peaks at frequencies close to proposed Doppler signals in at least two planetary systems, HIP24275 and HIP90988, calling for more long-term activity studies of giant stars.  
  Even disregarding these possible false-positives, extrapolation leads to a fraction of 25$\--$30\% of low-luminosity giant stars hosting planets.  We find the mass-function exponentially rises towards the lowest planetary masses, however there exists a $\sim$93\% probability that a second population of giant planets with minimum masses between 4-5~\mj\, is present, worlds that could have formed by the gravitational collapse of fragmenting proto-planetary disks.  Finally, our noise modelling reveals a lack of statistical evidence for the presence of correlated noise at these RV precision levels, meaning white noise models are favoured for such data sets.  However, different eccentricity priors can lead to significantly different results, advocating for model grid analyses like those applied here to be regularly performed.  By using our Bayesian analysis technique to better sample the posteriors, we are helping to extend the planetary mass parameter space to below 1~M$_J$, building the first vanguard of a new population of super-Saturns orbiting giant stars.    
  
  }

   \keywords{giant planet formation -- techniques: radial velocities -- Planet-star interactions}

   \maketitle
%

\section{Introduction}

The study of planetary systems orbiting intermediate mass stars, those with masses $\ge$ 1.5~M$_{\odot}$, or evolved stars has mainly been carried out using the radial-velocity (RV) method (e.g. \citealp{johnson07,niedzielski07,ppps1,jones11,ottoni22}).  This may be surprising, giving that large survey transit work from both the ground and space has recently given rise to an explosion of exoplanet detections.  However, the dominant orbital architecture of these systems, coupled with the observational strategies we employ to discover and characterise them, renders the RV method the most efficient tool in this work.  
First off, stars with an effective temperature (\teff) hotter than around 6500~K, present few spectral lines with which to perform RV analyses, and therefore those more massive stars on the main sequence inhibit precision RV surveys. Hence, to gain RV access to the planetary systems orbiting those stars, it is necessary to observe them when they have evolved off the main sequence.
At that point they cool down enough to present a rich forest of spectral lines that can be used to perform precision RV analysis.  
Furthermore, these studies have found a desert region between these stars and the first inner gas giant planet that orbits them, reaching out to around 0.5~AU (e.g. \citealp{Sato2008,Jones2013}).  This means that there are very few gas giant planets orbiting close enough to their host star to transit, and since the transit probability is inherently low, few transiting worlds have been detected orbiting both massive stars on the main sequence and their giant star counterparts. 
However, those that have been detected (e.g. \citealp{Lillo-Box2014,jones18,Saunders2025}) have unveiled interesting phenomena such as radius re-inflation \citep[e.g.][]{grunblatt16}, and represent an exciting benchmark sample that is rich for further study.

Gas giant planetary systems orbiting giant stars have been found to show some remarkable properties, both in the number density of planets and their orbital distribution.  Beyond the aforementioned lack of planets orbiting giant stars, it has also been shown that these planets are formed more often orbiting more metal-rich hosts (\citealp{johnson10b,reffert15}), similar to those orbiting main sequence (MS) stars (\citealp{fischer05,sousa11,jenkins13}).  The overall planet fraction increases with increasing stellar mass, until around 2~M$_{\odot}$ (\citealp{jones16}), and the fraction of giant planets orbiting evolved stars is high, around 10\% (\citealp{wolthoff22}), and up to $\sim$33\% for low-luminosity red giant branch (RGB) stars \citep{jones21}. Therefore, these stellar systems are rich hunting grounds to search for giant planets.

With a high abundance of giant planets orbiting field giant stars in the solar neighbourhood (up to $\sim$ 200 pc), we can study them in detail to better understand their orbital properties, which provides a window into the dynamic past of these planets, along with the formation histories of planets orbiting intermediate-mass stars on the MS.  Since the distribution of giant planets orbiting giant stars has been primarily mapped out using precision RV studies, the orbital parameters and physical characteristics like planetary masses, are heavily dependent on the RV models employed.  In the past decade, the RV community has moved towards searching the parameter space using more sophisticated tools, such as Markov chains and Bayesian statistics, or other tools like genetic algorithms, coupled with more complex correlated noise models like Gaussian Processes or Moving Averages (e.g. see \citealp{gregory07,tuomi09,feroz11,segransan11,jenkins14b,faria18} and refs therein). These approaches have led to a significantly improvement in the detection and extraction of genuine Doppler signals, particularly when combined with RV (quasi) periodic variability induced by stellar phenomena, such as magnetic activity and stellar pulsations.

The simultaneous inclusion of stellar activity noise models in the search for Doppler signals can provide strong gains when dealing with Doppler signal confidence.  In the past, more evolved K-type giant stars than those we study in this work, have shown (quasi) periodic signals that share Doppler-like properties.  Stars like Aldebaran, $\gamma$Draconis, and 42 Draconis exhibit long-period RV signals that could be interpreted as having planetary origins (i.e. \citealp{dollinger09,hatzes15,hatzes18}).  However, it seems these signals may have a stellar origin, with the exact physical phenomena still under debate (\citealp{reichert19,dollinger21}). More expansive noise modelling coupled with Bayesian statistics can help shed some light on such issues, and in the future, circumvent them. 

In this work we apply a Bayesian approach to search for new planets orbiting giant stars from a joint analysis of the RV timeseries data obtained by two giant star planet search programs, the EXoPlanets aRound Evolved StarS (EXPRESS; \citealt{jones11}) and the Pan-Pacific Planet Search (PPPS; \citealt{ppps1}) projects.  We also reevaluate additional systems with previously published planet candidates to check their validity.  We pay special attention to testing various models with different priors and correlated noise models, with the goal of arriving at the most appropriate statistical model to explain the data.  The manuscript is therefore organized as follows: the observations, data reduction, and RV measurements are presented in section \ref{sec2}. In section \ref{sec3} we present the host star properties. The RV analysis and Bayesian orbital fitting is presented in section \ref{sec4}.  The new discoveries and known planetary system tests are discussed in section \ref{sec5}.  We present the stellar activity analysis in section \ref{sec6}, with results from the general analysis of the population of known RV planets orbiting giant stars from this project being found in section \ref{sec7}.  Finally, some discussion points and the summary are found in section \ref{sec8}.

\section{Observational Data \label{sec2}}

High-resolution spectra from three separate instruments, (UCLES, CHIRON, and FEROS) were taken for the two new giant stars presented in this work. The observations for both of these targets were acquired as part of the EXPRESS and PPPS surveys, representing an amalgamation of data from two separate independent projects.  These two projects complement each other well, with samples that somewhat overlap, both having 37 stars in common, and a number of planetary systems have already been published using their combined data sets (e.g. \citealt{jones16,wittenmyer2016,wittenmyer17,jones21}). An additional 11 discovered planetary systems are also analyzed in this work, using RVs taken by the same three spectrographs, and used to better understand the population of planets orbiting low-luminosity giant stars.  The RV observations obtained with instruments are described in the three subsections below. 

\subsection{UCLES data} \label{sec:express_data}  

The PPPS operated from 2009-2015 as a collaboration between Australia, Caltech, and the National Astronomical Observatories of China.  It was a Southern Hemisphere extension of the established Lick \& Keck Observatory survey for planets orbiting northern ``retired A stars'' \citep[][e.g.]{johnson06, johnson07, johnson10b}.  All RV data from the survey were made available in \citet{ppps8}.  Data from the PPPS and EXPRESS were also used to compute occurrence rates of stars orbiting low-luminosity giant stars in \citet{wolthoff22}.  We obtained observations between 2009 and 2015 using the UCLES high-resolution spectrograph (\citealp{diego90}) at the 3.9 meter Anglo-Australian Telescope (AAT).  UCLES, now decommissioned, achieved a resolution of 45,000 with a one-arcsecond slit.  An iodine absorption cell provides wavelength calibration from 5000 to 6200\AA.  Precise RVs are determined using the iodine-cell technique as detailed in \citet{Butler1996}.  The photon-weighted midtime of each exposure is determined by an exposure meter.  All velocities are measured relative to the zeropoint defined by the iodine-free template observation.  The UCLES velocities are given in Tables~\ref{tab:rvs_HIP18606} and \ref{tab:rvs_HIP111909}.

\subsection{Chiron data} \label{sec:express_data}

We observed a total of $\sim$ 50 EXPRESS targets with the CHIRON \citep{chiron} high-resolution fiber-fed spectrograph mounted on the 1.5\,m telescope at the Cerro Tololo Inter-american Observatory (CTIO), in Chile. For each of these targets, we collected typically 10-15 individual spectra, between 2011 and 2023. For the observations we used the slicer mode, which leads to a resolving power of $\sim$ 80,000, and the I$_2$ cell in the light-path, and we adopted typical exposure times between 3 and 10 minutes, depending on the stellar magnitude. The data reduction was performed automatically by the CHIRON reduction pipeline \citep{Paredes2021} and the the final velocities were computed using the I$2$ cell method \citep{Butler1996}, with the pipeline described in \citep{Jones2017}. The resulting RVs are listed in Tables \ref{tab:rvs_HIP18606} and \ref{tab:rvs_HIP111909}.   

\subsection{FEROS data} \label{sec:express_data}

We observed the EXPRESS targets with the Fiber-fed Extended Range Optical Spectrograph (FEROS; \citealt{feros}) high-resolution spectrograph, mounted on the 2.2\,m telescope, at La Silla Observatory, between 2010 and 2024. FEROS is equipped with a secondary fibre that was illuminated by a simultaneous ThAr lamp, to correct for the nightly drift. We have obtained between 10 and 15 spectra per star, and for the planet candidates we have collected further data, typically reaching up to $\sim$20 epochs to characterize the planetary orbital parameters. The exposure times varied between $\sim$ 2 minutes for the brightest targets to 10 minutes for the faintest ones in the sample, leading to a S/N per extracted pixel $\gtrsim$ 100. The data were reduced with the CERES pipeline \citep{ceres}, that also computes the stellar RV, while the activity indicators were computed with the pipeline presented in \citet{Jones2017}.

\begin{table*}
\centering
\caption{HIP18606 Radial-velocity \& Spectral Diagnostic Measurements }
\begin{tabular}{lccccc}
\hline\hline
\vspace{-0.3cm} \\
BJD & RV & Error & BIS & $S$-index & Inst \\

[d] & [\ms] & [\ms] & [\ms] &  &  \\
\hline\\
2454866.94471  &  -10.24  & 1.89 & -- & -- & UCLES \\
2454868.90014  &   -7.83  & 1.45 & -- & -- & UCLES \\
2455139.15703  &    7.80  & 2.02 & -- & -- & UCLES \\
2455402.33690  &  -21.94  & 1.74 & -- & -- & UCLES \\
2455525.09656  &   -9.68  & 1.40 & -- & -- & UCLES \\
...\\ 
\vspace{-0.3cm} \\\hline\hline
\end{tabular}\\
This is only a short sample, the full table can be found in the online CDS platform.\\
\label{tab:rvs_HIP18606}
\end{table*}

\begin{table*}
\centering
\caption{HIP111909 Radial-velocity \& Spectral Diagnostic Measurements }
\begin{tabular}{lccccc}
\hline\hline
\vspace{-0.3cm} \\
BJD & RV & Error & BIS &  $S$-index & Inst \\

[d] & [\ms] & [\ms] & [\ms] & & \\
\hline\\
2455074.19065  &   -5.89  & 1.23 & -- & -- & UCLES \\
2455496.00896  &    0.00  & 2.04 & -- & -- & UCLES \\
2455879.95984  &  -31.05  & 2.31 & -- & -- & UCLES \\
2455905.91924  &  -17.60  & 2.28 & -- & -- & UCLES \\
2456052.28143  &  -10.10  & 2.68 & -- & -- & UCLES \\
...\\  
\vspace{-0.3cm} \\\hline\hline
\end{tabular}\\
This is only a short sample, the full table can be found in the online CDS platform.\\
\label{tab:rvs_HIP111909}
\end{table*}

\section{Stellar Modelling \label{sec3}}

For each of the stars in our sample, (the two new planetary systems we announce in this work and the 11 stars with previously detected planets), we employed a two-step approach to calculate the stellar parameters.  Firstly, we made use of the stellar parameter estimates from the Spectroscopic Parameters and atmosphEric ChemIstriEs of Stars (SPECIES; \citealp{soto18}) algorithm.  Although initially developed for use on main sequence dwarf stars, with a catalogue of over 580 dwarf star parameters presented in \citeauthor{soto18}, a SPECIES parameter catalogue for all giant stars in the EXPRESS project was also developed by \citet{soto21}.  This is the catalogue from which we draw the values used in this work, and therefore we refer the reader to those papers for the details of how SPECIES calculates the stellar parameters.  We do note however, that the $T_{eff}$, log$g$, and [Fe/H] are measured from equivalent width fitting of high-resolution stellar spectra, whereas additional bulk properties like mass and radius make use of the Isochrones package (\citealp{morton15}).  We also make use of the calculated equivalent evolutionary points (EEPs; see \citealp{dotter16}), which gives rise to the most likely position of the star in evolutionary space.  The EEPs allow us to discretely sample late-time stellar evolution, placing our stars either at the Terminal Age Main Sequence (TAMS), the Tip of the red giant branch (RGBTip), the Zero-age core helium burning (ZACHeB), or the Terminal-age core helium burning (TACHeB).

The second method we employ for calculating bulk stellar parameters is by using the ARIADNE code (\citealp{vines22}). ARIADNE is an open-source Python package designed to automatically fit the spectral energy distribution (SED) with several grids of atmospheric models in order to determine parameters such as $T_{eff}$, log$g$, and radius, among others.  ARIADNE operates as follows: firstly, it searches for publicly available photometric data on the star, then the SEDs are modelled using dynesty \citep{speagle20} with four different grids, the BT-Settl set \citep{allard12,castelli04,kurucz93}, and Phoenix v2 \citep{husser13}, and finally all models are averaged using the model probabilities as weights through a Bayesian Model Averaging approach, obtaining a final set of bulk parameters for each star.

As priors, we used the values derived from the SPECIES analysis for the $T_{eff}$, log$g$, and [Fe/H]. The extinction prior is a uniform distribution from zero to the maximum extinction in the line-of-sight from the SFD galactic dust maps \citep{schlegel98, Schlafly2011}. Finally, we take the distances from \cite{bailerjones21} as a prior for the distance.  The final calculated properties for all stars are shown in Table~\ref{tab:stellar_par}.

\begin{figure}
\includegraphics[angle=0,scale=0.45]{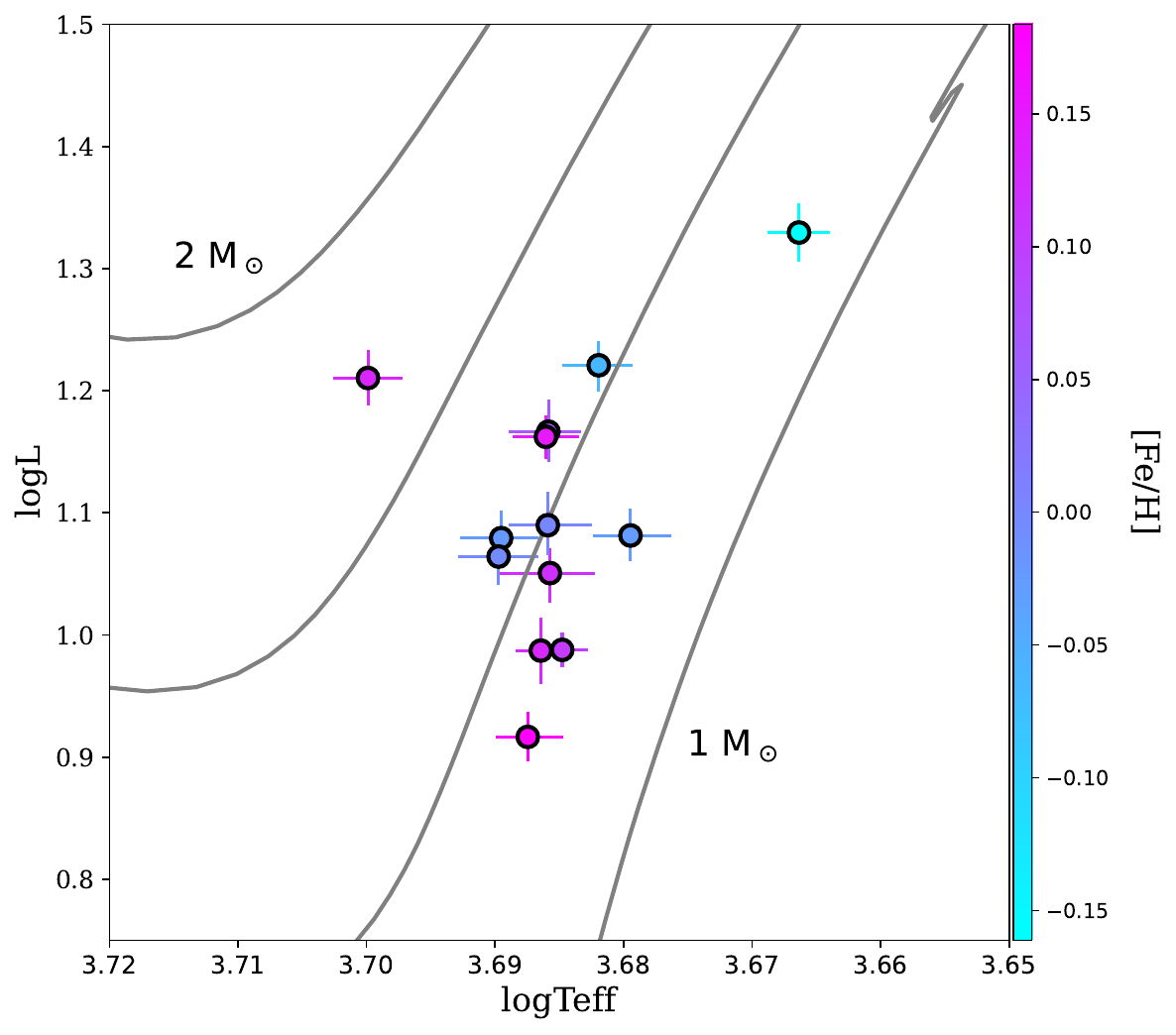}
\caption{HR diagram showing the position of all 13 target stars in this work (filled circles). The solid evolutionary curves were computed within the ARIADNE processing using MIST models, running between 1.0 \msun and 2.0 \msun\, and at a fixed solar metallicity.  The colours of each of the data points corresponds to the stellar metallicity, with the colour scale shown on the right hand side y-axis. }
\label{hrdiagram}
\end{figure}

\section{RV Modelling}
\label{sec4}

To analyse all RV data, we make use of our Exoplanet Mcmc Parallel tEmpering Radial velOcity fitteR (\emp) code version 1 (\citealp{vines23,pena25}).  \emp\, is an algorithm that is highly adapted to search for Keplerian signals in timeseries RV data sets.  The code employs Markov-chain Monte Carlo sampling, using the \emcee\, package (\citealp{ForemanMakey2017}) to generate posterior samples for our targets. We employ the parallel-tempering routine within \emcee, generating five chains with different temperatures, running from the `coldest' statistical chain, to progressively `hotter' chains that ensure the full parameter space has been well sampled.  Parallel-tempering is the process of raising the likelihood to some power ($\mathcal{L}^{\beta}$, where $\beta = 1/T$, with $T$ the `temperature' of the chain), such that narrow regions of high probability in the parameter space are broadened and damped across ever hotter chains.  When considering the swap of states between chains, this process allows the cold chain, (unmodified likelihood; $\beta=1$), to move more easily throughout the parameter space of interest, guarding against being captured in non-global maxima of the posterior.  In our case we set the short temperature ladder $\beta$ to approximately \{1.00, 0.49, 0.22, 0.11, 0.05\}.

\subsection{Bayesian Formulism}

Since \emp\, uses tempered Markov chains, we can make full use of Bayesian probabilities to manoeuvre the samplers to find and settle into the maximum of the posterior.  We can also make use of the posterior probabilities once we locate the maximum to calculate the statistical probability of the model, or how favoured it is by the observed data.  We can start with Bayes' Theorem: 

\begin{equation}
\label{eq:bma_1}
    p\left(\vec{\theta} | D, M\right) = \frac{p\left(D | \vec{\theta}, M\right) p\left(\vec{\theta} | M\right)}{p(D | M)}
\end{equation}

Where $D$ is the observed data, $\vec{\theta}$ is the parameter vector corresponding to model $M$, $p\left(D | \vec{\theta}, M\right)$ is the likelihood distribution, $p\left(\vec{\theta} | M\right)$ is the prior distribution, and the denominator $p(D | M_i) = \int p\left(D | \vec{\theta_i}, M_i\right) p\left(\vec{\theta_i} | M_i\right)d\vec{\theta_i}$ is the marginalised likelihood of model $M_i$ with parameters $\theta_i$, which is also referred to as the Bayesian evidence, the quantity required such that the posterior integrates to unity over the parameter range.

The global model that we employ in this work is made up of a combination of seven terms and three nuisance terms.  The form of the expression is as follows:

\begin{equation}\label{eq:bayesmodel}
r_{i,j} = \gamma_{j} + \dot{\gamma_{j}t_i} + f_{k}(t_{i}) + \epsilon_{i,j} +\eta_{j,h} + \zeta_{j,q}~,
\end{equation}

with $r_{i,j}$ being the modelled RV for the $i^{th}$ observation in the timeseries and $j^{th}$ instrument, $\dot{\gamma}$ is a linear acceleration parameter per instrument as a function of time $t$ to take care of any significant long period trends in the data, and $\gamma_{j}$ takes care of any offsets between the $j^{th}$ instruments.  $\epsilon_{i,j}$ is our white noise term that is the quadratic sum of the instrumental uncertainties, modelled as a zero mean normal distribution with a standard deviation equal to the measurement uncertainty (i.e. $\mathcal{N}(0, \sigma_i^2)$), and the excess stellar jitter noise, represented by a zero mean normal distribution with the standard deviation representing the jitter (i.e. $\mathcal{N}(0, \theta_{jit}^2)$).  The  $\eta_{j,h}$ term represents the linear correlation functions for each instrument $j$ that we introduce to model any additional noise that can be tracked by the measured $h$ activity indicators, such as the chromospheric $S$-index or the bisector inverse slope (BIS). Note that the FWHM of the FEROS CCFs did not show any correlations or periodicities for any targets and so were not included.  The form of these linear terms are given by $\sum\limits_{l} c_{l,i}l_{i,j}$, with $c$ being the correlation coefficient for each activity index $l$, normalised by its respective RMS. 
 In reality, each of these stars has been preselected to represent likely RV stable stars (see \citealp{jones11}).  For example, applying the expected RV pulsation amplitude scaling relations (\citealp{kjeldsen95,samadi07,kjeldsen11}), only three of the stars have pulsation amplitudes expected to be $\ge$10~\ms, with the largest predicted value being $\sim$20~\ms.  The estimated values as shown in Table~\ref{tab:stellar_par}. 

The $\zeta_{j,q}$ term relates to the correlated noise part, which we model using a moving average (MA) with exponential smoothing, given by 

\begin{equation}\label{eq:ma}
\zeta_{j,q} = \sum\limits_{q} \phi_{j,q} exp\left( \frac{-|t_{i-q} - t_i|}{\tau_{j,q}} \right)m_{i-q,j}
\end{equation}

where $q$ is the order of the MA, which in this work is fixed to unity to represent a first-order MA model.  $\tau$ is the exponential smoothing parameter, the decay timescale of the correlation for a given instrument and MA order, and $m$ represents the residuals of the model fit to the data.  The final term in the global model expression is $f_{k}(t_{i})$, which represents the Keplerian component that is given by
\begin{equation}
 f_{k} (t_{i}) = \sum_{n=1}^{k} K_{n} [\, \text{cos}( \omega_{n}+ \nu_{n}(t_{i})) + e_{n} \text{cos}(\omega_{n}) ]~,
 \end{equation}
 
 which is a function that describes a $k$-Keplerian model with $K_{n}$ being the velocity semi-amplitude, $\omega_{n}$ is the longitude of pericenter, $\nu_{n}$ is the true anomaly and $e_{n}$ is the eccentricity. $\nu_{n}$ is also a function of the orbital period and the mean anomaly $M_{0,n}$, representing the phase ($\Theta$) of the model function, a parameter we fit for.

By application of our model within the MCMC algorithm, at each step we must evaluate the probability of the proposal density, or the proposed new values for the parameter vector $\vec{\theta}$, and determine if the increase in probability warrants a selected movement of the chain within the hyper-dimensional space we are probing.  Therefore, we require both a description of the likelihood function $p\left(D | \vec{\theta}, M\right)$ and prior probability $p\left(\vec{\theta} | M\right)$.

\subsection{Likelihood Function}

Within \emp\, we make use of a Gaussian likelihood function, which takes the form:

\begin{equation}\label{eq:like}
    \mathcal{L}(\theta)=p(D|\vec{\theta},M)=\frac{1}{\sqrt{\det{\bf{E}}}}exp\bigg[-\frac{1}{2} \sum\limits_{uv} (d_u - r_u)[\bf{E}^{-1} \it]_{uv}(d_v-r_v)\bigg]
\end{equation}

Where now we have introduced $\bf{E}^{-1}$ as the inverse of the covariance matrix, implemented to take care of any correlated uncertainties we have in the data ($d$).  For clarity we have set the subscripts $u$ and $v$ to be the positions within the matrix, columns and rows respectively.  For fully independent uncertainties, the matrix itself is diagonal and this part of the expression simplifies to the $\chi^2$, being replaced with the term $\sum_u\frac{(d_u - r_u)^2}{\sigma_u^2}$, where $r_u = r_{i,j}$, the different data points $i$ and instruments $j$.  Given that the likelihood is translated into log-space, we omit the constant term of $2\pi^{N/2}$ in the denominator outside of the exponential.

With the likelihood constructed as shown in Eqn~\ref{eq:like}, we require a kernel to fill out the positions of the covariance matrix, or model the correlations.  In our case, and as shown in Eqn~\ref{eq:ma}, we model the correlations using a first-order MA, meaning we only deal with the correlations two steps either side of the diagonal, and the remaining positions in the matrix are set to zero.  This serves to perform local smoothing of the data, whilst saving large fractions of computing time due to simpler matrix inversions.

\subsection{Prior Choice}

In order to efficiently sample the RV data, we set the priors for our Bayesian probabilities to be similar to those we have utilised extensively in the past when searching for small signals in noisy data.  Orbital period and RV amplitude are the two dominant parameters that define the success of the chain samplings to locate the maximum of the posterior.  We set a wide uniform prior on both of these parameters, constrained to be three times the maximum time baseline of the RV data for the period, and three times the  maximum spread in the RV measurements for the amplitude.  The eccentricity is the only Keplerian parameter that is non-uniform, with it being set to a folded Gaussian function, centered at zero and either with a standard deviation of 0.1 (e$_{TC}$) or 0.3 (e$_{C}$), depending on the models.  This effectively prioritises more circular orbits, which tend to be common outcomes of the planet formation and evolution process for giant star systems, yet it is flexible enough that planets on highly eccentric orbits are still allowed if the data strongly argue for their inclusion.  In any case, we do also test model sets with the eccentricity fixed to zero, representing purely circular orbital solutions (e$_{F}$).  The angular parameters for the Keplerian model are both uniform and set to the full extent between 0$\--$2$\pi$ radians.  We tend to find that angular parameters are not very well constrained in many cases, particularly with circular, or close to circular orbits, and therefore flat priors are most appropriate here.

For the noise parameters, we again utilise mostly uniform prior bounds, such that we allow the most flexibility across the samplings, with one exception again, the jitter parameter.  The offsets between the data sets ($\gamma$) are bounded to be between zero and the maximum of the RV measurement spread.  The MA coefficients ($\phi$) have a uniform prior between minus unity to unity, along with the linear acceleration parameter ($\dot{\gamma}$), and the MA timescale ($\tau$) is set to a fixed value of five days, which was arrived at after extensive testing across short and long timescales.  Finally, the jitter parameter ($\theta_{jit}$) maintains the normal distribution form as has been previously used with \emp, yet we change the location of the mean to be set to 5~\ms, since these giant stars are expected to have significant levels of {\bf p-modes-induced} related noise in RV data, at this level or more.  We therefore also set the standard deviation to be 5~\ms, allowing significantly higher values to be well sampled throughout the processing.  All prior values and formats are listed in Table~\ref{tab:prior}.

\section{New Planetary System Discoveries and Known Planet Tests}
\label{sec5}

We analyse the data employing various noise model approaches, assessing the model outcomes and best fit probabilities when including or omitting MA correlated noise models, and/or linear activity correlation terms with the BIS measurements and S-activity indices from the FEROS spectra.  Therefore, we run models with only white noise components (WNO models), a mixture of individual correlated noise components (BLCO or MA), combined linear activity components (BSLCO), or full correlated noise components (BLCMA and BSLCMA), giving rise to six separate model runs.  We also run for fixed eccentricity and with the eccentricity as a constrained parameter, employing three different eccentricity priors for each of these models.  The eccentricity priors are listed in Table~\ref{tab:prior}, with the 'Tightly Constrained Eccentricity' referring to the e$_{TC}$ prior, the 'Constrained Eccentricity' the e$_{C}$ prior, and the 'Fixed Eccentricity' the e$_F$ prior.  We then compare the Bayes Factors (BFs, parameterized as $\Delta$BICs) between each of these models internally, (the same noise or eccentricity model but different numbers of Keplerian components), and externally, (comparing each best fit model for different noise or eccentricity components), and we report the model that best describes the data.  

\subsection{HIP18606}

We model the 54 RVs for HIP18606 without including any linear activity correlations, nor any MA correlated noise components in the first instance, meaning we are applying a pure white noise model approach at this time.  We find a single statistically significant signal (BF $> 5$ or $>$150x more probable than the next best model) with a period and amplitude of 674.94$^{+6.43}_{-6.28}$~days and  16.07$^{+1.12}_{-1.18}$~\ms, respectively, for the e$_{F}$ WNO model, which turns out to be the best fit (see Table~\ref{tab:18606bfs} for the BF values for each model).  Including the BIS and S-index linear correlations without the MA components (BLCO \& BSLCO models) also clearly detects the Doppler signal, and including the MA components in the model (MA, BLCMA, and BSLCMA) consistently find similar signals.  We can see that the BLCO models have BIC values around 4-5 higher than the WNO models, whilst including the S-indices increases the BIC by a further 3-4, and then adding additional correlated components, (the MA models), increases the BIC by a further seven or more after that.  The fact that the MA models are the least favoured models in general, tells us that there is a lack of noise correlations for this data set, and so the simplest models better describe the velocity distribution for HIP18606.  No models across the three model sets found evidence for a second signal in the data.  

\begin{figure}
\includegraphics[angle=0,scale=0.35]{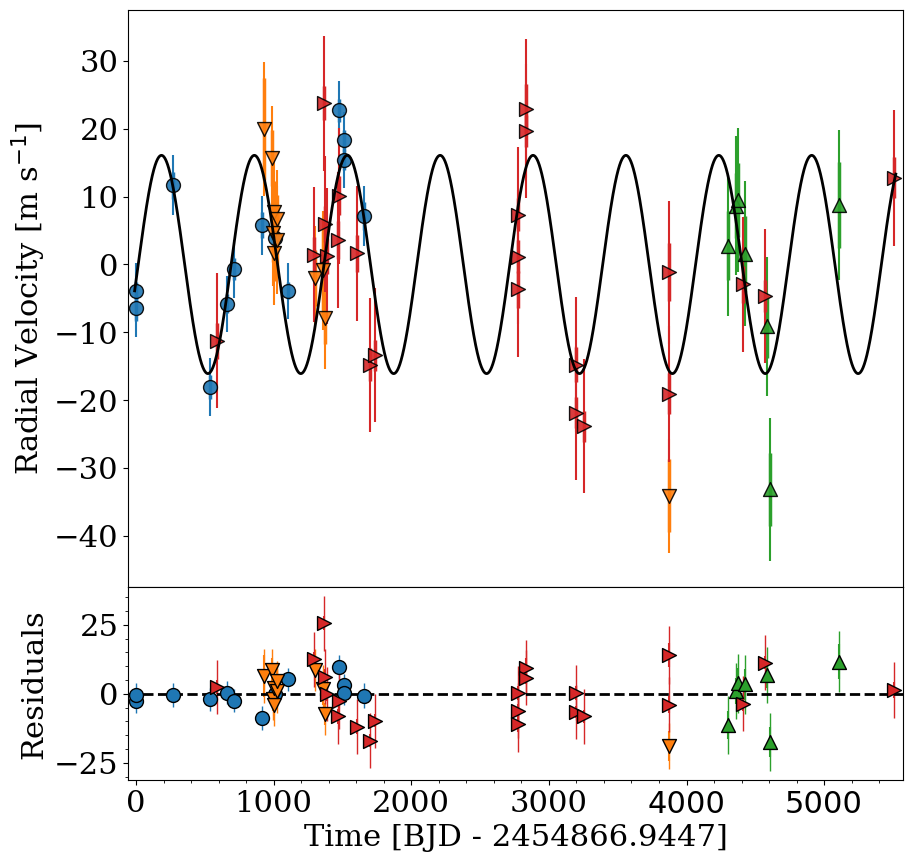}
\includegraphics[angle=0,scale=0.35]{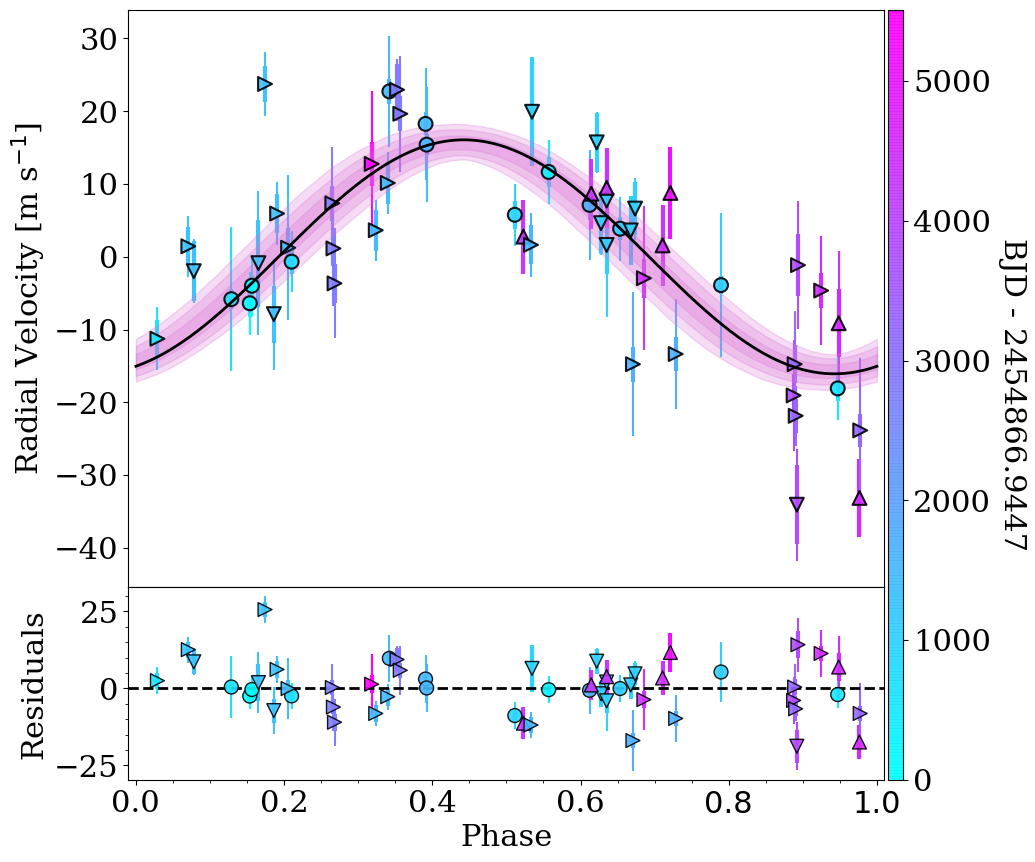}
\caption{In the top panel we show the RV timeseries for data taken with UCLES (blue circles), FEROS (red tilted triangles), and the pre- (orange triangles) and post-pandemic Chiron data (green triangles) for HIP18606.  The best-fit Keplerian model is represented by the black curve.  The lower panel shows the RVs phase folded to the period of the detected planet candidate.  The symbols represent the same data, however the points are coloured depending on their observing date, with the colour bar shown on the right of the plot, and with the pink contours around the best-fit Keplerian model in black delimiting the 1,2, and 3$\sigma$ spread taken from the Markov chains.  Both plots relate to the overall most probable model, the fixed eccentricity model with a white noise model component only.}
\label{fig:18606}
\end{figure}

Although the ensemble analysis is all in agreement with the existence of a single Doppler signal in the data, the solutions can be very different depending on which model set, or which type of eccentricity prior we employ.  In this case, the models within the fixed eccentricity (e$_F$) model set are highly favoured over the models that include eccentricity as a constrained parameter of the model (e$_{TC}$ \& e$_C$).  For the WNO models, the e$_F$ model set provides a BIC that is $\sim$8 lower than these other models.  This trend is mirrored by the other models within each of these model sets, indicating that circular solutions are the most likely configurations that can explain the current data.  Given that the e$_F$ models are less complex than the other model sets, the strong penalty on model complexity within the BIC could be biasing towards the selection of these circular models. However, interestingly, without correcting for the Chiron RV offset that appeared due to the full observatory shutdown throughout the recent global pandemic, the same signal is detected yet an eccentric solution is heavily favoured.  This highlights the need to continuously understand instrument stability when properly modelling RV data sets, and also that the model selection will favour more complex eccentric solutions if the data argue for them.  We rely on the calculated statistics and model complexity here to determine the best solution, and the bold e$_F$ WNO model is significantly favoured over all other models.  When compared to its closest rival, the BLCO model within the same e$_F$ model set, the BF is found to be 4.5, which means a probability of 90$\times$ more in favour of the WNO model.  The final full RV model and phase folded RV curve are shown in Fig.~\ref{fig:18606}, and the parameter estimates are found in Table~\ref{tab:18606}.

Since the search for additional companions in the RVs did not yield any confirmed candidates, we can therefore confirm the existence of a gas giant planet with a separation of just over 1.5~AU.  No statistically significant linear trend is apparent in the data, meaning there is no current evidence in favour of any more massive companions at much larger separations than the few thousand days time baseline of the data.  

\subsection{HIP111909}

In the case of HIP111909 when we fit the 52 RVs with all models, we find two statistically significant signals in all of them.  In addition, for all models we find a 3rd signal in the data, yet this signal only has the same period in the e$_{TC}$ and e$_{C}$ model sets, and a different period in the e$_F$ model set.  In fact, for a number of models we find a 4th signal that was statistically detected, yet the period of these were dependent on the model set assumed.  For example, for the e$_C$ set the BLCO and MA models found signals with periods of $\sim$435~days, whereas the fixed eccentricity set regularly found a fourth signal in the white noise only and bisector correlations only sets with a period of $\sim$782~days.  Since we did not find any agreement here for the 4th signal, we decided not to consider these as possible planets at this time, yet they are likely spurious noise related to poor model fitting for the k3 solutions we pick-up (see Table.~\ref{tab:111909bfs}).  In the end, we decided to focus on the fact that all models across all sets detected three statistically significant signals in the data, and we discuss below which of these we can consider as detected planet candidates at this time.

\subsubsection{Two Planet Solution Only}

Although we find excellent agreement for both signal parameters across all models for this data set, we find a range of probabilities (BICs: 457 \-- 497).  Following a similar pattern to the case of HIP18606, the e$_{TC}$ model set provides slightly lower probability model fits than the e$_{C}$ model set, indicating models with slightly more flexibility should be favoured, yet the best fits are found to be for the circular models.  We shall see this to be a recurring theme when looking at the rest of the stars we have analysed in this work, and hence eccentricity prior choice is important when searching for the 'true' model parameter values, even when the model choice is the same (i.e. a truncated Gaussian in this case, only the standard deviation is altered).

Another recurring theme we shall see is that the most probable set of models were found to be those without MA components for the e$_C$ and e$_F$ sets, with the overall best model the e$_F$ WNO model including two Keplerians (BIC=457.6).  In general for this dataset, adding the MA correlated noise components only served to significantly increase the BIC values, even though the same two signals were detected.  The best fit planetary model finds two signals with periods of 487.08$^{+3.81}_{-3.63}$ and 893.63$^{+14.89}_{-16.03}$~days, and RV semi-amplitudes of 25.57$^{+2.03}_{-1.52}$ and 13.86$^{+1.25}_{-1.14}$~\ms, for planet candidates HIP111909b and c, respectively.

\begin{figure}
\includegraphics[angle=0,scale=0.32]{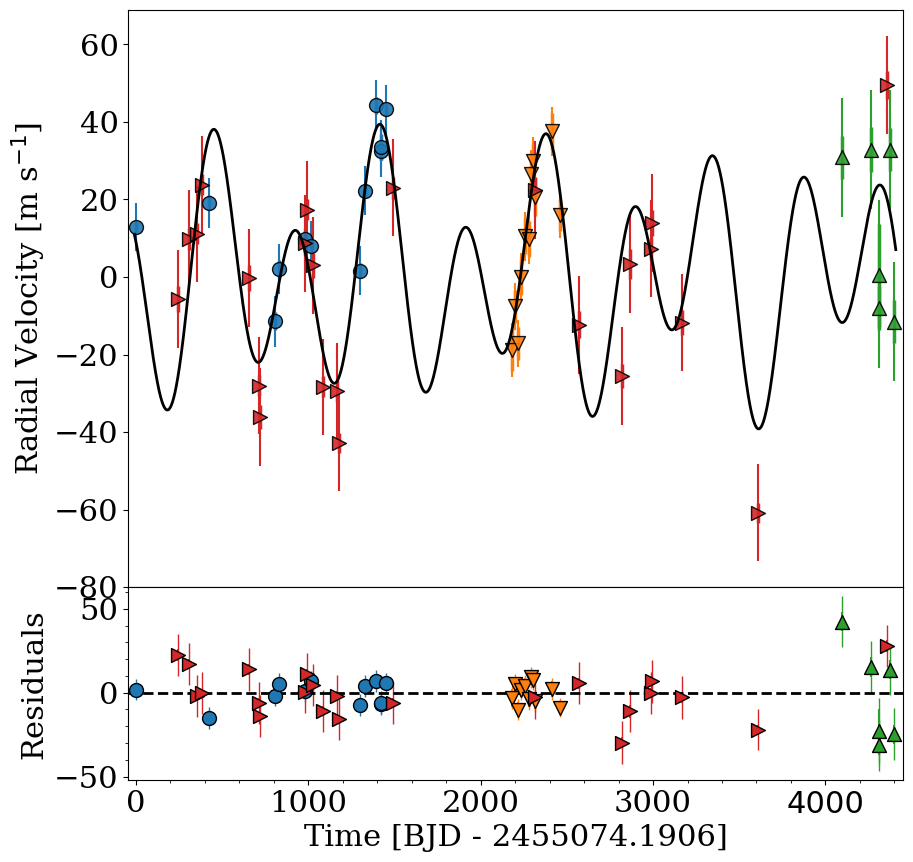}
\includegraphics[angle=0,scale=0.32]{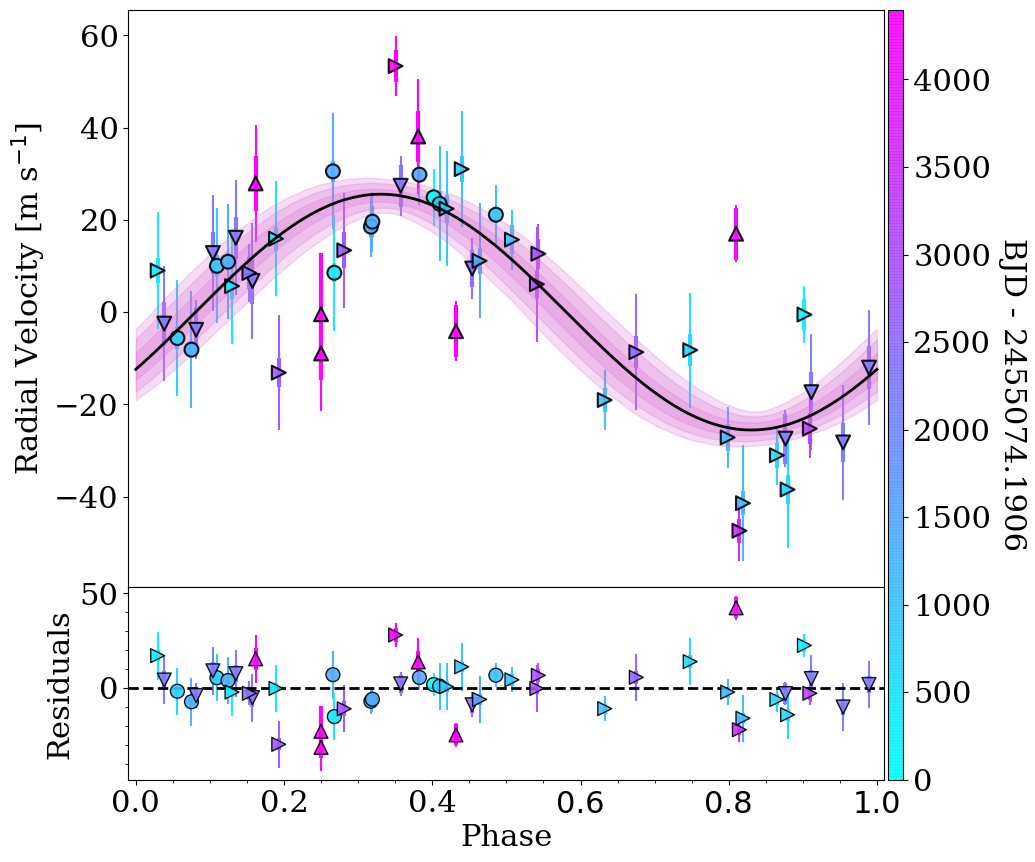}
\includegraphics[angle=0,scale=0.32]{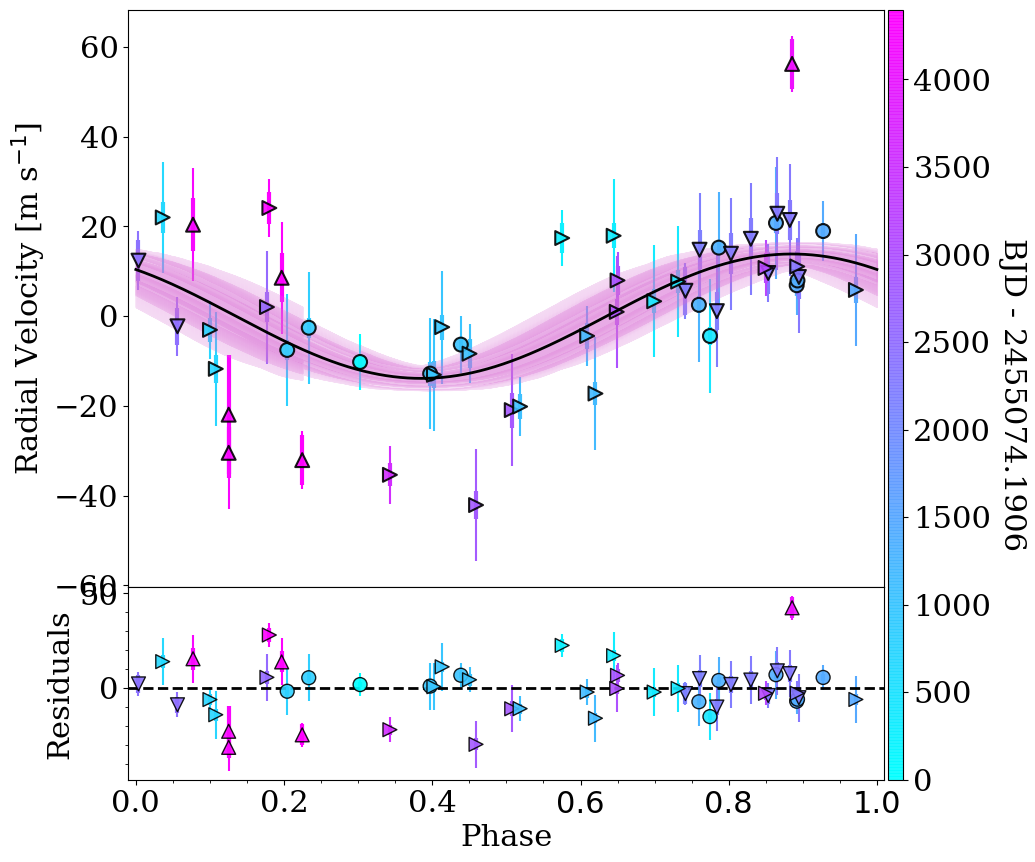}
\caption{The top panel of the upper plot shows the full timeseries RV data, including data from the UCLES (blue circles), FEROS (red tilted triangles), and Chiron pre- (orange triangles) and post-pandemic (green triangles), along with the joint best fit Keplerian model (black curve) for HIP111909.  The lower panel here shows the residuals to this model fit.  The center and lower plots shows the phase folded signals detected by \emp, along with the best fit models shown in black, and the 1,2,3$\sigma$ uncertainty bounds (pink shaded regions).  The lower panels highlight the residuals to the model fits.}
\label{fig:111909ecc}
\end{figure}

The resulting solution we report here gives rise to the discovery of two giant worlds with minimum masses of 1.21$\pm$0.10 and 0.81$\pm$0.08~\mj, for planets b and c respectively.  Their respective orbital periods are also intriguing, since the period ratio is found to be 1.84, very close to the 5:3 period ratio.  In fact, they are just over 3$\sigma$ away from this ratio, which hints at the possibility that these Jovian worlds were in or around the 5:3 mean motion resonance in the past, and were then possibly knocked out of resonance as the star evolved towards the RGB.  Interestingly, dynamical tests provide hints that the planets are currently still in resonance, motivating further detailed dynamical studies to reveal if such a scenario is possible.  We show the phase folded RVs with best fit Keplerians, along with the RV timeseries and best model, in Fig.~\ref{fig:111909ecc}, and the values for the signals drawn from this model are shown in Table.~\ref{tab:111909}.  

When looking at the excess noise, jitter, for the best fit solution, we find that the jitter is significantly larger for the post-pandemic Chiron data set, than when compared to the data taken prior to the observatory shutdown.  Although this may seem drastic and could point to a change in the star's activity status, or an instrumental issue with the Chiron spectrograph after reopening the observatory, we only have six RVs in the second Chiron set, and therefore it is most likely that this is just an issue of low number statistics.

\subsubsection{Third Detected Signal}

As mentioned, EMPEROR detected a 3rd statistically significant signal in all model sets and model runs.  However, as the period of the 3rd of signal was not constant over all model sets, we decided not to report either of these as a true Doppler signal at this time, even though at least one of them may be.  For the e$_{TC}$ and e$_C$ model sets, all models agree with the third signal having a period and amplitude of $\sim$17.5~days and $\sim$9~\ms, respectively.  For the e$_F$ model set, all models found a 3rd significant signal with a period and amplitude of $\sim$41.5~days and $\sim$10~\ms, respectively.  The amplitudes of the 3rd signal in all models agrees well, yet the e$_F$ model set consistently finds a 3rd signal that is nearly 2.5$\times$ larger than both the e$_{TC}$ and e$_C$ model sets, meaning there is no global agreement in the period of the 3rd signal as yet.  A by-eye inspection of the phase folded RV plots for both signals may indicate that the 17.5~day one looks cleaner, however the BIC is significantly better for the WNO e$_F$ model fit.  We caution that if planets b and c are actually in resonance, any third signal could arise due to evolution of the orbital angles and would therefore require a full Newtonian model analysis to confirm its reality.  Since the RV data does not statistically argue for any linear trend at this time, we are therefore left to conclude that HIP111909 is orbited by at least two gas giant planets, with orbital periods relatively close to a 2:1 period ratio.

\subsection{Known Planet Tests and Additional Candidates}

In addition to the newly discovered planetary systems, we also tested our method on known systems to confirm this independent method also finds them, whilst searching for new signals.  The 11 systems we studied were previously discovered using data from the UCLES, Chiron, and FEROS instruments, and we include here new spectroscopic data not used in the original papers. We report the BIC values from 18 different model tests per star in Appendix~\ref{appendixb}.  We also show the updated system parameters in Table~\ref{tab:confirm} and the RVs, both full timeseries and phase folded, along with the best-fit Keplerian models in Appendix~\ref{appendixc} (\href{url}{https://zenodo.org/records/17024692}). 

Overall we find strong statistical evidence for the existence of these signals, as expected due to the reported high RV amplitude level of each (see \citealp{johnson10,jones15a,jones15b,wittenmyer15,jones16,wittenmyer2016,wittenmyer17,jones21}).  However, our modelling tends to favour more circular orbits for a number of these systems where the eccentricity had previously been found to be statistically non-zero.  Moreover, we also detect four new signals in this analysis, indicating the possible presence of additional planets orbiting the stars HIP8541, HIP67851, HIP75092, and HIP90988 (we elaborate more on these below).  Finally, we confirm the linear trends for HIP74890 \citep{jones16} and HIP90988 \citep{jones21}, and we detect statistically significant (beyond 99\%) RV trends in HIP67851, HIP75092 and HIP95124, providing strong evidence for additional long-period companions in these systems.  In addition to giant planets orbiting within a few AUs, low-luminosity giant stars also appear to frequently host companions on much wider orbits.

\section{Stellar activity analysis \label{sec6}}

\subsection{Spectral Index Analyses}

To allow filtering of stellar activity signals that could impact the RV measurements, we extract a number of spectral indices from the high-resolution spectral timeseries for each star.  The indices we choose to make use of measure either the flux variations in spectral lines of interest, or line shape variations in the form of asymmetries or width changes.  For the flux variation indices we make use of the classical chromospheric $S$-index (\citealp{duncan91}), and both the H$\alpha$ and Na~\sc i\rm\, D3 line indices (\citealp{santos10}).  For the spectral line shape variations we choose to include the bisector inverse slope (BIS; \citealp{QuelozEtal2001aaHD166435}) and the full width at half maximum (FWHM; e.g. \citealp{AngladaEscudeEtal2016natProxCenb}) of the CCF used to measure the RVs.  These indices were used in three ways to study the activity, where 1) we analyse the indices using generalized Lomb-Scargle Periodograms (\citealp{zechmeister09}), 2) Keplerian fitting of the indices using standard manual cosine and Keplerian model fitting, and 3) by adding the indices into our EMPEROR Keplerian fitting prescription to model them simultaneously with the RVs, aiming to understanding their impact on the RVs directly.  All GLS periodograms can be found in Appendix~\ref{appendixa} (\href{url}{https://zenodo.org/records/17024692}), and therefore here we only discuss the four stars where there are potential activity signals that may relate to signals present in the measured RVs.

\begin{figure}
\hspace{-0.45cm}
\includegraphics[angle=0,scale=0.7]{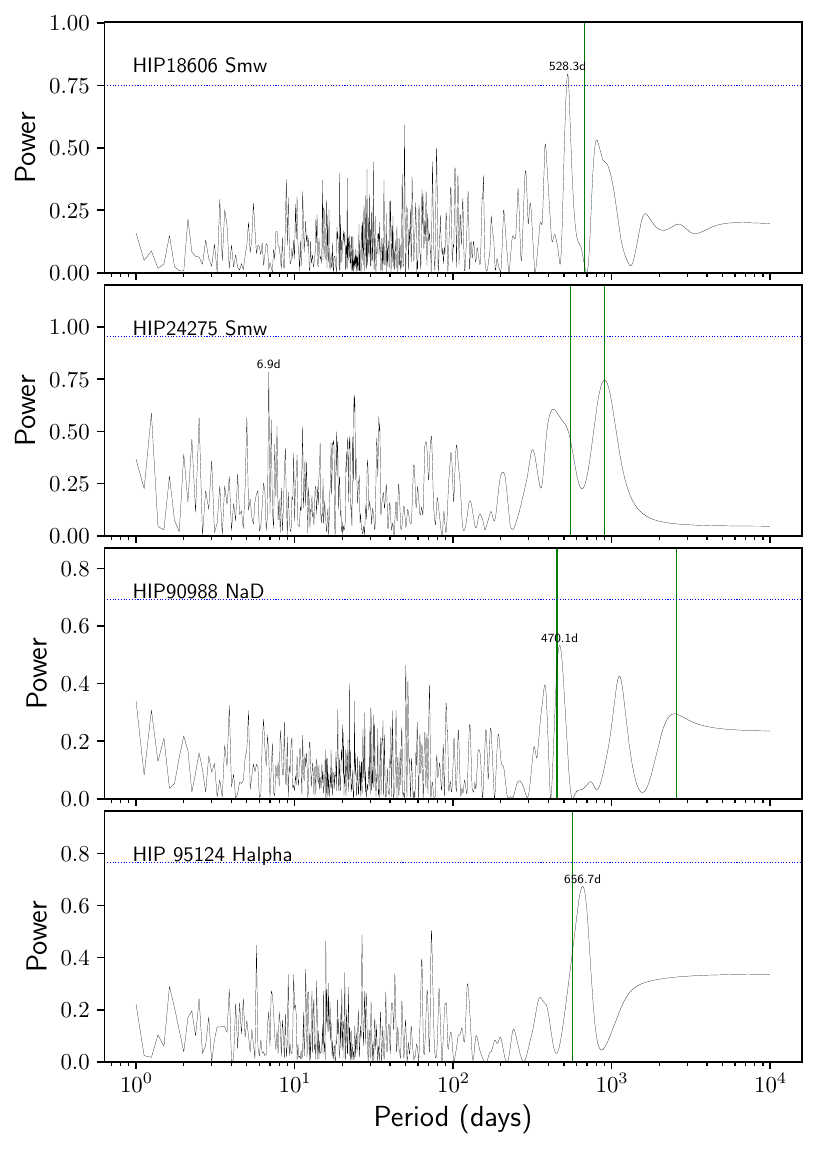}
\caption{Periodograms for four stellar activity indices that show peak powers close to a candidate planet signal detected for four of the stars in this sample.  From top to bottom we show the $S$-activity index periodograms for HIP18606 (top) and HIP24275 (second top), the Na\sc i\rm~ index for HIP90988 (second bottom), and the H$\alpha$ index for HIP95124 (bottom).  The green vertical lines mark the positions of candidate planets and the horizontal blue dotted lines show the 0.1\% false alarm probabilities.  The periods of the strongest peaks are also labelled in each panel.}
\label{fig:act}
\end{figure}

No stars in this sample showed any significant periodogram peaks in the line asymmetry indices that were found close to a planet candidate signal frequency.  The activity indicators on the other hand did yield one such signal, and three others with weak peak powers that are close to planetary periods.  HIP18606, HIP24275, HIP90988, and HIP95124 all have detected signals that could be flagged as potential false positives.  \linebreak

\paragraph{HIP18606}

In Fig.~\ref{fig:act} we show the four periodograms for these stars, ranked from top to bottom in name ascending order.  The top panel shows the periodogram of the $S$-index timeseries for HIP18606, and this is the only statistically significant signal that crosses the 0.1\% false alarm probability (FAP) limit and is relatively close to the detected planet candidate period.  However, we can see that the peak frequency detected for this signal is $\sim$146~days different from that of the planet candidate.  We also ran an EMPEROR model of this timeseries, (as in \citealp{rubenstein25}), and both the $S$-index signal and planet signal are statistically separated at the 17.9$\sigma$ level.  Therefore, we can safely rule out this activity signal as the culprit for the detected Doppler candidate signal in the RV timeseries. \linebreak

\paragraph{HIP24275}

The second panel in Fig.~\ref{fig:act} also shows the periodogram of the $S$-indices, but this time for the star HIP24275.  No peaks are statistically significant in this analysis, with the largest being found at a frequency of 6.9~days.  However, a close second is the period that a manual fit to the timeseries gives as 912~days, which closely matches the candidate Doppler signal at 905~days.  The \emp\ results find a signal with a period of $883.8^{+33.9}_{-38.9}$~days, however the model is rejected as it narrowly misses the statistical significance condition previously mentioned. The proximity of the \emp\ signal in the $S$-indices to the RV signal might suggest that it is related to activity, though more data is needed to acquire any statistically significant result since there are only 11 epochs of data. 

As mentioned, both of these planets were originally published in \citet{wittenmyer2016} and our analysis above including activity modelling also did not remove this signal, therefore we cannot definitively show that activity is driving the lower frequency RV signal, but there exists the possibility that this is the case.  Moreover, most of the models we ran (Appendix~\ref{appendixb}) actually found three planet candidate signals, and even if the proposed planet c is found to be a false positive, additional data may yield a more definitive detection of a second planet at a period of around 290~days.  No strong activity peaks are found in any of the periodograms for this star close to 290~days.

The first signal for this RV timeseries relating to the planet candidate b at a period of 550~days, appears to be sandwiched between the peak at 912~days and another around 430~days.  The shorter period of these two signals also seems to be heavily asymmetric, which might mean there is a third peak hidden here that better matches the period of the planet b Doppler candidate signal, but we could not confirm that at this time, meaning more spectral data is needed to test if this is the case. \linebreak

\paragraph{HIP90988}

The third panel in Fig.~\ref{fig:act} shows the periodogram of the sodium (Na) D line timeseries for the star HIP90988, again with no statistically significant signal as yet.  However, the strongest peak closely matches the period of the previously detected planet b signal at $\sim$452~days (\citealp{jones21}).  In fact, the current data gives a peak period of 470~days from the periodogram, with a best fit modelled signal period of $470.5^{+6.3}_{-7.3}$~days, which gives rise to a difference of only 2.4$\sigma$ between the best fit RV Doppler signal period and this activity-induced signal.  Therefore, although both periods seem to be separated, they are not statistically distinct.  Testing the planet candidate here with further data in the future is crucial, to ensure that the activity frequency is sufficiently well constrained and does not further approach that of the proposed planet.  We also note that the second planet candidate at longer period, newly detected in this work, falls close to a small bump in the periodogram around 2560~days.  In this case, the bump maximum is significantly lower than many other peaks at different frequencies, and can not be interpreted as a potential activity-induced signal in the RV timeseries at this stage.  Again, more spectral data will help to elucidate this issue. \linebreak

\paragraph{HIP95124}

Finally, the fourth panel in Fig.~\ref{fig:act} that relates to the periodogram of the H$\alpha$ timeseries for the star HIP95124, also shows the strongest statistical signal appearing close to the frequency of the detected planet b.  The activity signal peak is found to be 656~days, whereas the planet signal is at 564~days, nearly 100~days separated.  Additionally, when we run a best fit model to the activity data we find an activity period that is more than 100~days separated from the planet candidate Doppler signal, and therefore it is unlikely to be the origin of the Doppler candidate.  Furthermore, \emp\ does not detect any significant signals in the H$\alpha$ data, hence we conclude that HIP95124b is likely a genuine planet orbiting this star.

\subsection{ASAS Photometric Analysis}

To further investigate whether stellar phenomena could be the source of the observed RV variations (e.g., \citealt{boisse11}), we analyzed the All Sky Automated Survey (ASAS; \citealt{asas}) $V$-band photometry of all giant stars in this work.  In order to ensure we are making use of the highest quality data sets for these tests, we only used A- and B-grade ASAS data, which are known to be relatively free of various controllable sources of noise. Due to evidence of stability issues in the early photometric data, we calculated the mean and standard deviation of all data \textit{after JD 2452300}. Using the statistics from the more stable data, we then applied a 3\,$\sigma$ clipping filter to the entire baseline of the data, to remove serious outliers.  This type of selection can also have the effect of removing data that were taken when the star was in another epoch of its activity cycle that is far from the norm, like moments of extreme inactivity or extreme activity.  However, this is also a desired effect since we are aiming to search for longer-term stable signals like that from the star's rotational frequency.  In the top and middle panels of all plots in Appendix~\ref{appendixd} (\href{url}{https://zenodo.org/records/17024692}) we show the raw and cleaned data for each of our targets, and we can see that the majority of the removed outliers come from early in each timeseries, when the photometric quality was less stable.  

\begin{figure}
\includegraphics[angle=0,scale=0.35]{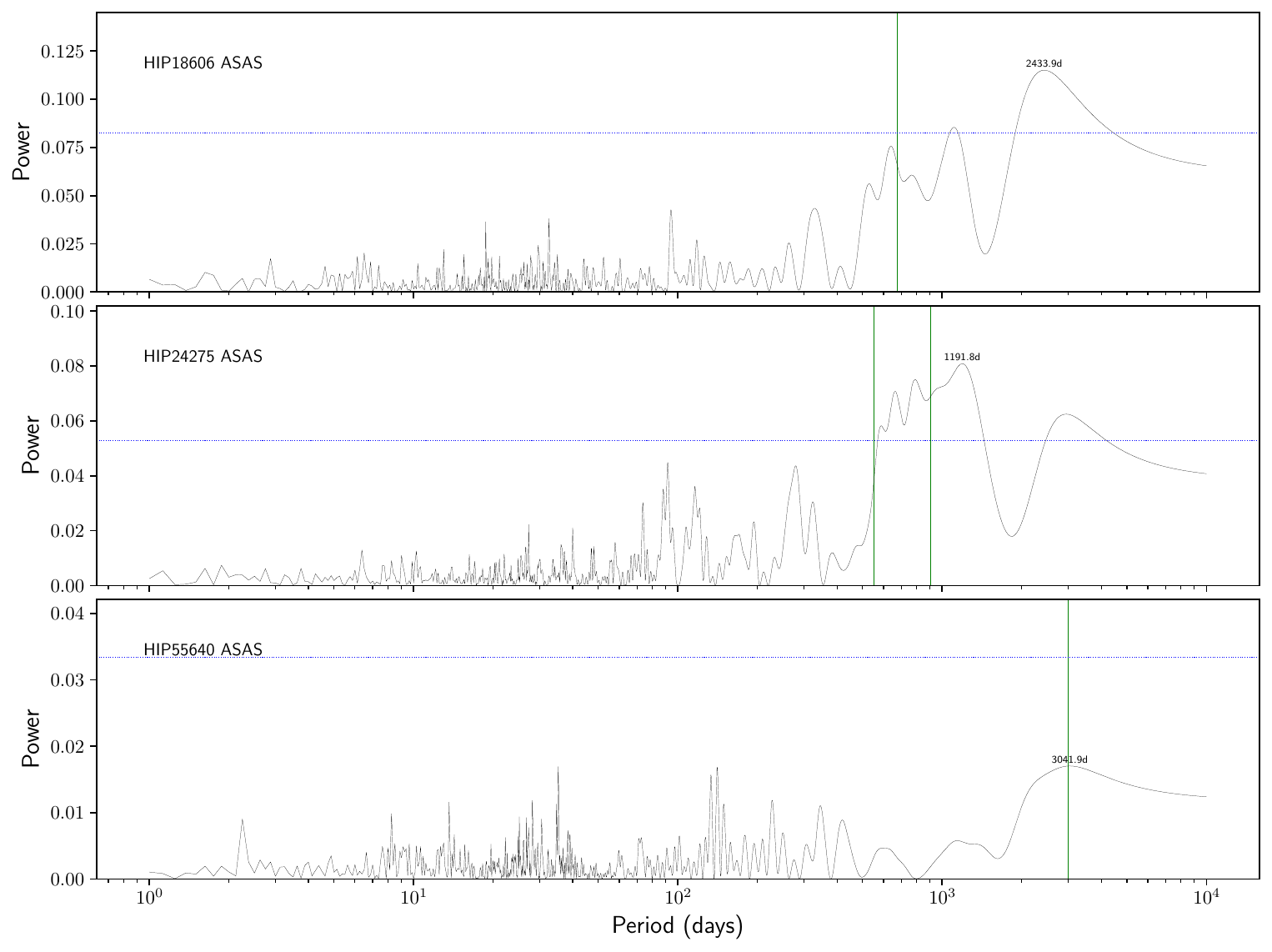}
\caption{GLS periodograms of the cleaned ASAS timeseries photometry for the stars HIP18606, HIP24275, and HIP56640.  The vertical lines mark the positions of the detected planet candidates in the RVs for each star, whereas the horizontal blue dashed lines represent the 0.1\% FAP thresholds.  The name of each target is also highlighted in the top left corner of each respective panel.}
\label{fig:joint_asas_gls}
\end{figure}

After cleaning the ASAS data we again passed each data set through a GLS periodogram analysis to search for any stable and statistically significant signals present in the timeseries, and we show all of these in the bottom panels of the figures in Appendix~\ref{appendixd} (\href{url}{https://zenodo.org/records/17024692}).  Only six of the sets showed statistically significant signals present in the data, with the rest showing nothing of interest.  A few of these are attributed to signals arising due to the overall baseline of these discretely sampled timeseries, a common issue with periodogram analyses, whereas others seem to be genuine photometric frequencies found to be present on these stars at the time of the observations.  However, what we were mostly interested in once again were the signals arising close to the frequencies of the planet candidates detected by the EMPEROR analysis.  We show the three cases of interest that require discussion in Fig.~\ref{fig:joint_asas_gls}.

The top panel in Fig.~\ref{fig:joint_asas_gls} shows the results for HIP18606, which also was highlighted above in the spectral activity analysis of the $S$-index values.  In fact, both the ASAS photometry and $S$-indices showcase a broadly similar pattern around the planet candidate frequency, with signal strength around 500$\--$600~days.  Here the photometry peak is closer to the RV signal period, although still not quite overlapping, with a period of 640.07~d.  It must be noted that this power peak is not the maximum strength in the periodogram, but the 3rd highest frequency peak, below the 0.1\% FAP.  It also doesn't quite match the peak frequency found in the $S$-indices, which was found at a significantly lower period, whereas the flanking peak at lower strength does match the $S$-index frequency.  This may indicate that the $S$-indices are tracing stellar surface spot movements, at least at these frequencies.  

The center panel shows the results for HIP24275, with the outer planet candidate RV signal falling in an area of increased power in the ASAS GLS periodogram.  Indeed, this data set also shows a weak peak in $S$-indices at the period of the proposed RV Doppler signal, and with the addition of this cluster of significant peaks surrounding the detected RV frequency, this casts serious doubt on the reality of HIP24275c, likely indicating that this RV signal arises due to stellar surface phenomena and not an orbiting planet.  The actual maximum is found to be at 1191~d, however there are four statistically significant peaks surrounding the RV signal in this region, showing that the star presents an activity cycle, likely varying or multiple, at frequencies that are causing our algorithms to detect a signal in the RVs.

The bottom panel shows the GLS periodogram for the star HIP56640, and although there are no statistically significant peaks at any frequency across the period space of interest, the largest peak corresponds to the period of the candidate Doppler signal in the RVs for this star.  

\subsection{Hipparcos Photometric Analysis}

\begin{figure}
\includegraphics[angle=0,scale=0.35]{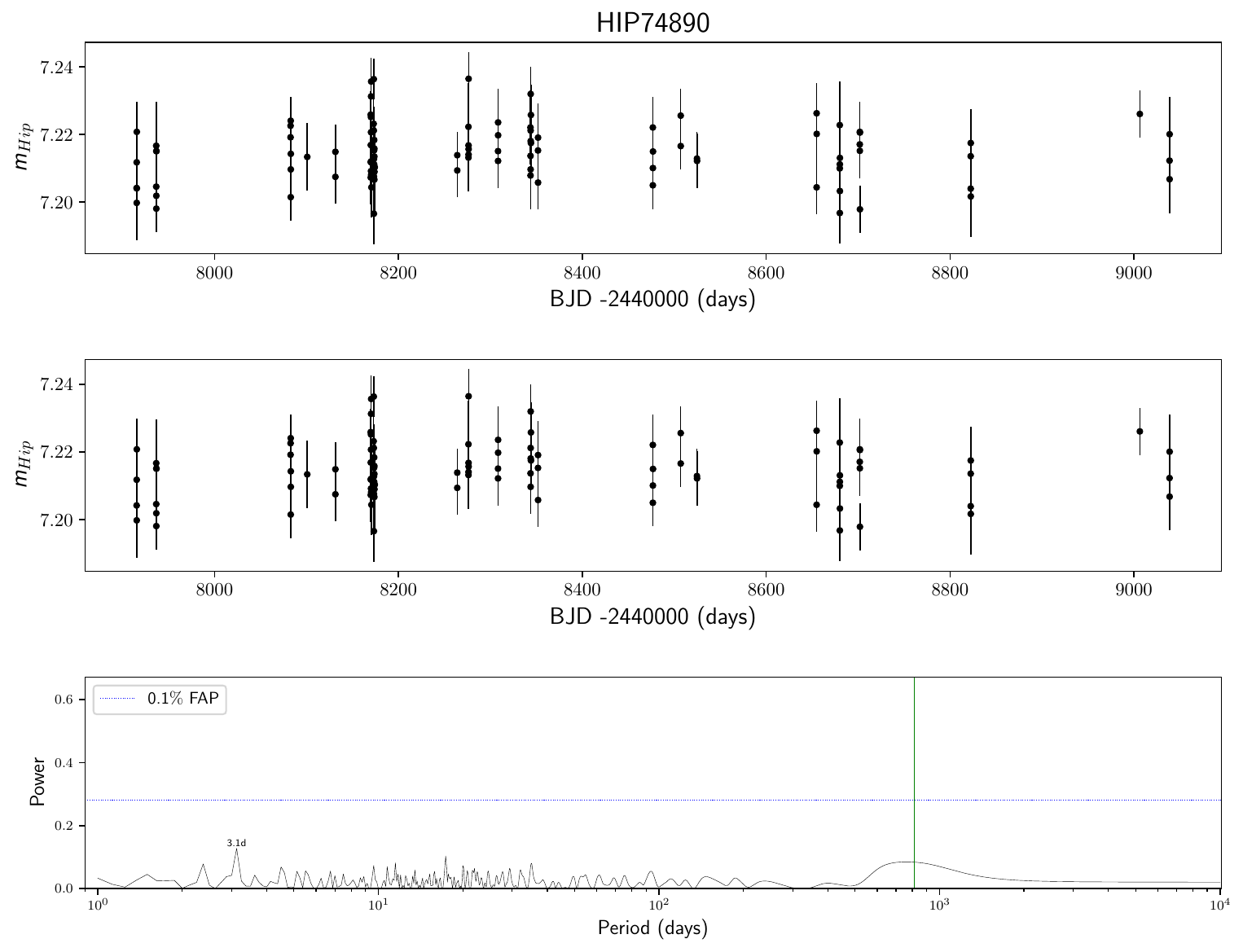}
\caption{Hipparcos photometric data for HIP\,74890 is shown in the top panel.  The middle panel shows the cleaned Hipparcos photometry for HIP\,74890.  The generalized Lomb-Scargle periodogram of the photometry is shown in the bottom panel. The horizontal line corresponds to the 0.1\% (dotted blue) FAP significance level computed via 5000 bootstrap iterations on the photometry time series. The vertical green line represents the periods of the planets.}
\label{hipp:74890}
\end{figure}

We also used the Hipparcos photometry \citep{leeuwen07} to search for potential periodic signals in the data that could match the orbital period of the published and newly proposed planets. For this, we included only data with quality flag equal to 0 or 1, and we analysed the data in the same manner as the ASAS data, applying our GLS techniques to look for significant periodicities in the photometry.  None were found for any of our sampled stars.  In fact, only one periodogram showcased a small local peak close to a planetary candidate signal, that of HIP\,74890.  In Fig.~\ref{hipp:74890} we show the data and periodogram, along with the position of the proposed planet.  Visually there does appear overlap, yet statistically this does not appear to be the case.  The photometric signal is found to peak at 762d, whereas the Keplerian period is 812d, significantly well separated given the uncertainties.  We conclude therefore that Hipparcos photometry also does not rule out any proposed Doppler signals in our sample.  All additional raw and cleaned Hipparcos photometeric timeseries, along with the GLS analyses, are shown in Appendix~\ref{appendixe} (\href{url}{https://zenodo.org/records/17024692}).

\section{General Analysis Results}
\label{sec7}

\paragraph{Noise Properties}

\begin{figure}
\includegraphics[angle=0,scale=0.64]{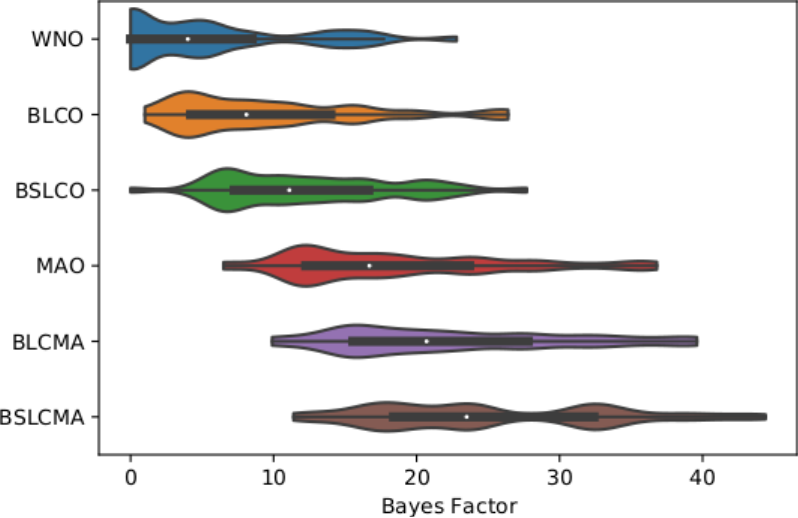}
\includegraphics[angle=0,scale=0.64]{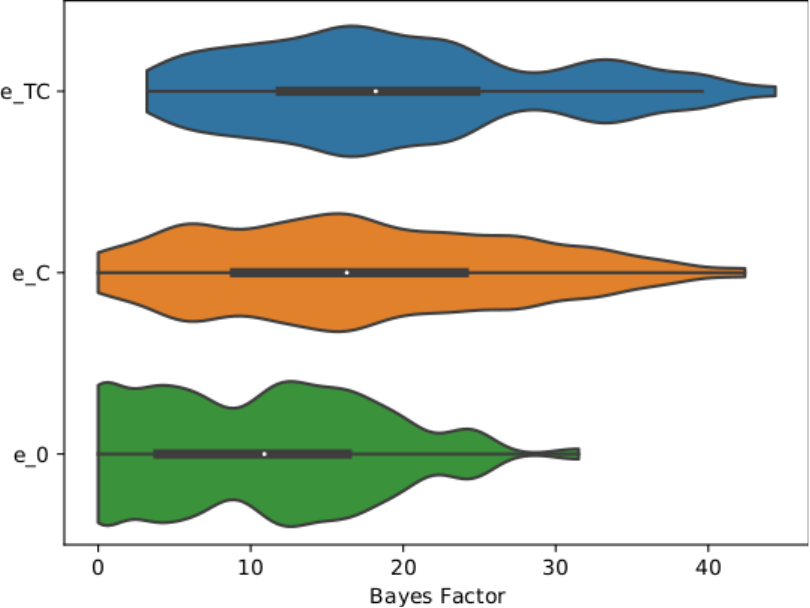}
\caption{The top panel shows a violin plot that represents the distributions of the Bayes Factors, ($\Delta$BICs), when compared to the best fit for all models applied to each of the four realisations of the noise tested in this work.  The blue curve represents the WNO models, the orange curve highlights the BLCO models, the green curve shows the MAO models, and the red curve the BLC+MA models.  The lower panel shows another violin plot that highlights the distributions of the same Bayes Factors, but this time separated by the three eccentricity priors that were tested in this work.  The blue distribution is for the e$_{TC}$ prior, the orange distribution is for the e$_C$ prior, and the green distribution is for the e$_F$ prior.}
\label{fig:violin_noise}
\end{figure}

One of the general results from the analysis of these data sets is that the inclusion of a correlated noise model such as a moving average, is highly disfavoured when compared to more simple models that assume pure white noise only, or even simple linear correlations between the RVs and spectral line stability diagnostics like bisector inverse slopes or the chromospheric $S$ index.  In the top panel of Fig.~\ref{fig:violin_noise} we can see the distribution of the BFs for each of the noise models, following a violin plot format, and we find that the mean and spread increases as a function of model complexity.  In addition, the density of these probabilities is highly condensed towards small values for the WNO models when compared to the other noise models, peaking towards BFs of zero.  This illustrates well that for these low-luminosity giant stars, employing models that assume gaussian distributed noise only is generally a good approximation of the data.  The other noise model distributions do not peak towards zero, but the position of their peaks also increase in BF with model complexity.  Therefore, making use of more advanced analysis techniques for giant stars beyond periodograms, like those employed here, should focus more on properly sampling the posterior space to 1) constrain better the uncertainties and uniqueness of detected signals and 2) to detect new signals close to the intrinsic noise level of the data, rather than to model high-frequency noise correlations following frameworks that have successfully been applied to dwarf stars on the main sequence (e.g. \citealp{tuomi2013b}; \citealp{jenkins13}; \citealp{AngladaEscudeEtal2016natProxCenb}).

The high intrinsic noise level of these giant stars, commonly dubbed \emph{jitter}, is one explanation for the lack of correlations in the measured RV noise.  Additionally, the timescales of correlated noise may not be conducive to the RV sampling employed in surveys like EXPRESS.  These stars generally rotate slowly, (\vsini$\sim$1-4~km/s), meaning they have long rotational periods and low expected activity induced signals from surface features like spots, signals that are difficult to sample in a high cadence RV survey.  The sample has been selected to have $B\--V$ values below 1.2 to ensure RV stability (see \citealp{hekker06}), and this also doubles as a crude estimation of the coronal dividing line from (\citealp{haisch99}), meaning they are expected to host hot coronas.  However, hosting a hot corona can be independent of the activity level (\citealp{hunsch96}), and therefore these stars are still expected to exhibit relatively low photometric variability from effects like rotating star spots (\citealp{henry2000}).  However, \citeauthor{henry2000} suggest that the variability that these stars do show is driven primarily by non-radial pulsations, and from their Table~1 the periods are generally less than one month, (low-luminosity giants like these here are expected to have periods of less than one day), with the main outlier showing a period of 240~days.  Therefore, stars like these with deep convective envelopes and large, extended atmospheres, could maintain pulsation frequencies that again work against a high cadence RV survey.  In the end, these types of RV surveys are biasing against potential correlated noise, aiding in the discovery of genuine Doppler signals from orbiting planets.

\paragraph{Eccentricity Distribution}

In the lower panel of Fig.~\ref{fig:violin_noise}, we show the same violin plot format, but for the BF values separated by the three different eccentricity priors we employed here.  First of all, we can see that the tightly constrained prior, e$_{TC}$, had a minimum that was higher than both the constrained prior, e$_C$, and the circular fixed prior, e$_F$, values, indicating it was the poorest choice overall.  However, it has a more condensed appearance than the e$_F$ models, showcasing a lower spread across the BFs, in close agreement with the distribution of the e$_C$ models.  This suggests that these models did not produce descriptions of the data as poor as those using the e$_F$ prior, and in general agreement with the e$_C$ prior models.  We may have expected that the e$_{TC}$ and e$_C$ prior model sets would produce fairly similar results, and this indeed appears to be the case, with the e$_C$ model sets producing a similar overall distribution, but with slightly lower spread.  Interestingly, the e$_F$ probabilities peak towards zero BF, highlighting that circular models generally explain the data better than models that allow for some non-zero eccentricity, even if that eccentricity is statistically similar to zero.  Yet the distribution spread is generally broader around the lower BF values, it does not taper down towards the zero value, indicating that the circular models fit the data well but can be less constraining than the other two prior model sets.

\begin{figure}
\includegraphics[angle=0,scale=0.35]{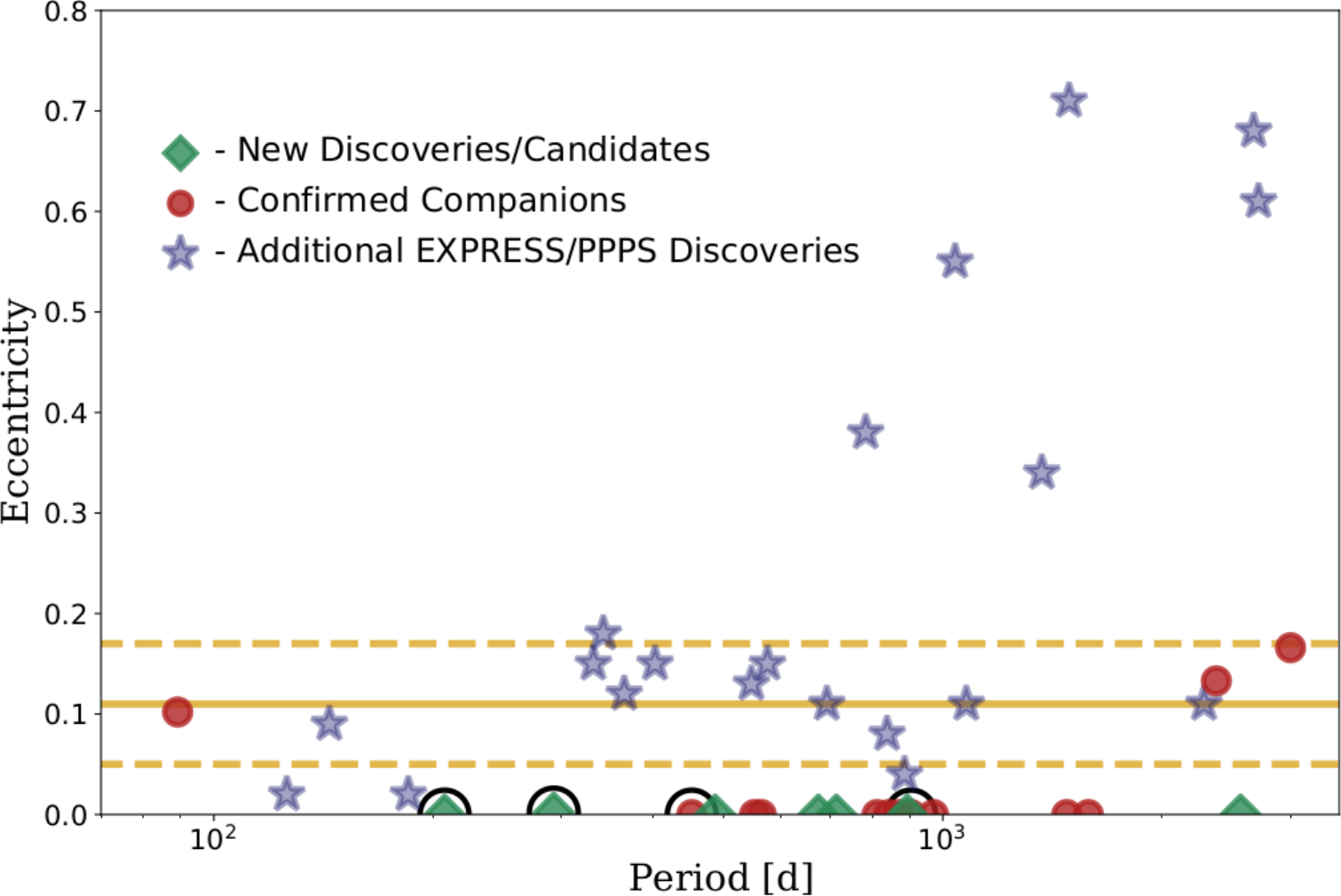}
\caption{The distribution of eccentricities of the giant planets within the EXPRESS and PPPS overlapping samples, as a function of their orbital period.  The filled green diamonds represent the newly discovered planets in this work, (including the possible candidates HIP24275d and HIP75092c for reference), with the red circles highlighting the additional confirmed planets, and the purple filled stars all other planets from these projects not included here.  The solid and dashed horizontal yellow lines mark the mean eccentricity of the 12 previously detected planets we include in this work and their associated 1$\sigma$ uncertainty bounds, respectively.  The ringed detections are those of HIP24275b \& d, HIP75092c, and HIP90988b, highlighted for the reasons discussed in the text.}
\label{fig:per_ecc}
\end{figure}

Taking these results at face value, it appears that the gas giant planet population orbiting these types of giant stars, favour circular orbits in general.  However, just like Sun-like dwarf stars on the main sequence, they exhibit a population of massive companions whose orbits are significantly different from zero (e.g. \citealp{wittenmyer2017}; \citealp{jones18}; \citealp{bergmann21}), the so called eccentric giants, even though in general their planets tend to be found on more circular orbits than the Sun-like population (\citealp{jones2014}).  In Fig.~\ref{fig:per_ecc} we show the period-eccentricity distribution for the planets in the EXPRESS and PPPS samples, highlighting the populations of planets on circular orbits and the population of eccentric giants.  In addition, we also find that our Bayesian formulism argues for the planets with eccentricities in close agreement with zero, (i.e. e$<$0.2), to actually have circular configurations.  Of all the 11 systems we analysed, the confirmed planets (red circles) are almost all exclusively on circular orbits, with only the one multi-planet system HIP67851 showing significant non-circular orbits.  We also show in the plot the mean and standard deviation ($\mu$=0.11, $\sigma$=0.06) of the published eccentricities for these planets, which is the regime where the majority of planets are likely to reside on even more circular orbits, and we see they are less than two standard deviations away from zero.  We can test the other planets in our samples (purple stars) in future works to see how many indeed favour circular orbits, but for the time being, these results are broadly consistent with the conclusions in \citet{bowler2020}, since the sub-population of long-period planets with eccentricities greater than 0.3 in this figure, contain a number of massive companions with a mean minimum mass of 8.5$\pm$7.2~\mj, with all three companions with minimum masses greater than 10~\mj\, being found in this parameter space.

\paragraph{Period - Minimum Mass Distribution}

\begin{figure}
\hspace{-0.3cm}
\includegraphics[angle=0,scale=0.36]{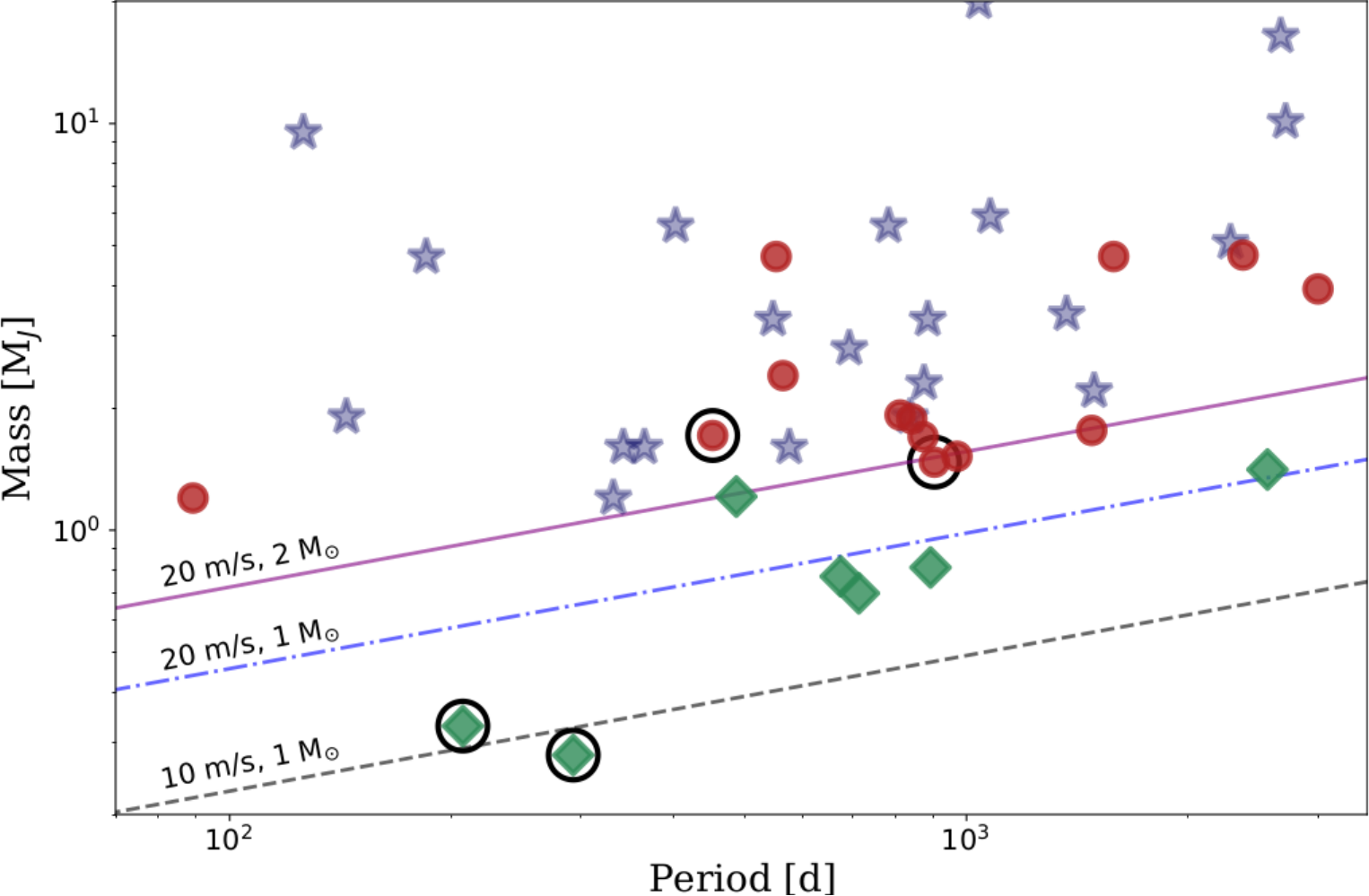}
\vspace{0cm}
\caption{The distribution of giant planet minimum masses as a function of orbital period discovered orbiting low-luminosity giant stars from the overlapping sample in EXPRESS and PPPS projects.  As before, the filled green diamonds represent the newly discovered planets in this work, with the red circles highlighting the additional confirmed planets, and the purple filled stars all other planets from these projects not included here. For reference, the 10~\ms\, and 20~\ms\, RV sensitivity limits for a 1~\msun\, star are shown by the dashed and dot-dashed curves, respectively, and the 20~\ms\, limit for a 2~\msun\, star is shown by the solid curve.  The ringed detections are those of HIP24275b \& d, HIP75092c, and HIP90988b, highlighted for the reasons discussed in the text.}
\label{fig:per_mass}
\end{figure}

All of the confirmed planets in this work have minimum masses in the range 0.3$\--$5.0 Jupiter-masses, and orbital periods between 90$\--$2800 days.  To gain a better understanding of how our new discoveries and updated parameter estimates fit in the global picture of giant planetary systems, we show the distribution of planetary minimum masses as a function of orbital period in Fig.~\ref{fig:per_mass}.  We can see that the previously discovered planets that are represented by the red circles, all fall significantly above the 20~\ms\, sensitivity limit for a solar mass star, with the lowest mass objects straddling the 2~\msun\, limit.  This population of so-called 'low hanging fruit', are expected to be the first set of planets uncovered, since their Doppler amplitudes are much higher than the instrumental sensitivity limits and the intrinsic stellar jitter (see \citealp{hekker06}).  However, using our new method of better sampling the posterior parameter space, including more apt models for the RV data and including some additional RVs, the newly discovered planets marked by green diamonds fall mostly below this sensitivity limit.  Actually, at the shortest orbital periods they reach down to the 10~\ms\, RV limit for a solar-mass star, however we must caution here that the two possible lowest mass planet candidates are ringed to indicate that there is some evidence from the activity index analyses that these are related to stellar activity.  Although the activity analysis for HIP\,8541 did not yield any conclusive results, we note that the RV amplitude relating to planet candidate c is below that expected for the pulsations of this star from the empirical scaling relations. 

HIP\,67851 represents an interesting case since recent results from \citet{fontanet25} using data acquired with the Coralie spectrograph, (refer to that work for details on the observations), give rise to a significantly longer orbital period for planet c than is found using our independent data set analysis (see Fig.~\ref{fig:67851}).  This tension arises due to the poor sampling of the outer signal in our data, which is being constrained mostly by the two final RV points in the timeseries.  Indeed, by including the Coralie data in our analysis, we find a much longer period planet, in good agreement with their published solution.  

In addition to the newly constrained planet c solution, our analysis gives rise to a new planet candidate, HIP\,67851d.  The signal is found to have a frequency between those of planets b and c, existing outside of the inner 2:1 period ratio of planet c (see Fig.~\ref{fig:67851c}).  Given there appears no indications of stellar activity affecting the RVs at this frequency, and none of our correlated noise models remove the signal, there is a strong probability that this arises from a real third gas giant orbiting this star.  However, by performing a dynamical analysis of the system using the parameters shown in Table~\ref{tab:confirm} for the Coralie included data, we find that the system likely could not have such eccentric orbits.  For the masses found, where all three are above a Jupiter-mass, only circular orbits at these periods are found to be stable.  Anything non-circular causes rapid orbital evolution and the ejection of planet c.  Statistically, circular orbits indeed describe the system well, however more eccentric orbits are preferred.  We do note however that this could be affected by the process used in this work to constrain the previously Keplerian solution(s) once sampling for an additional one.  The fact the eccentricities are constrained may present a small bias against finding better circular solutions, and so future work should be done to test this.

As for HIP\,75092c, this would be the lowest mass planet yet discovered by RV analysis orbiting a giant star, hosting a minimum mass of 0.33$^{+0.04}_{-0.03}$~M$_J$.  We draw caution here with the interpretation of this result, since 1) only when fixing the eccentricity prior to zero does this signal reach statistical significance, 2) there are two competing solutions at very similar, but distinct, periods, and 3) not all the models in the e$_F$ model set significantly detect this new signal.  We do note that all the models in all model sets do arrive at the same k2 signal, however the BFs never reach the $\Delta BF$ threshold of five to claim a statistical detection in those other cases.  Therefore, more RV data is required to move forward with confidently calling this a newly detected low-mass planet.  In any case, this population seems to trace out the real sensitivity limits of our data, as they rise in minimum mass towards the longest orbital periods we have sampled.  This visually highlights the impact that detailed Bayesian modelling of the data can provide, opening up a new population of lower-mass planets orbiting giant stars without the need to wait many more years for significantly higher volumes of RV measurements to be acquired.

For the time being, it is difficult to draw any conclusions about the period-minimum mass distribution of giant planets orbiting these giant stars, particularly for this new population of lower-mass planets we have uncovered here.  Given the limitations imposed by the RV sensitivity biases in the data, we can say that there does not appear to be a significant correlation between orbital period and planetary minimum mass, beyond the known lack of gas giants on short period orbits ($<$100~days or so).  This new population of super-Saturns are too few to provide any first statements about their nature, even though they are appearing to populate a similar part of the parameter space as their more massive counterparts.  In order to properly study the overall population, it is necessary to apply our Bayesian methodology to a much larger sample of these stars, and then statistically test to what level the RV biases are affecting our results, similar to the work of \citet{tuomi14} for lower-mass stars.

\paragraph{Mass Function}

\begin{figure}
\includegraphics[angle=0,scale=0.32]{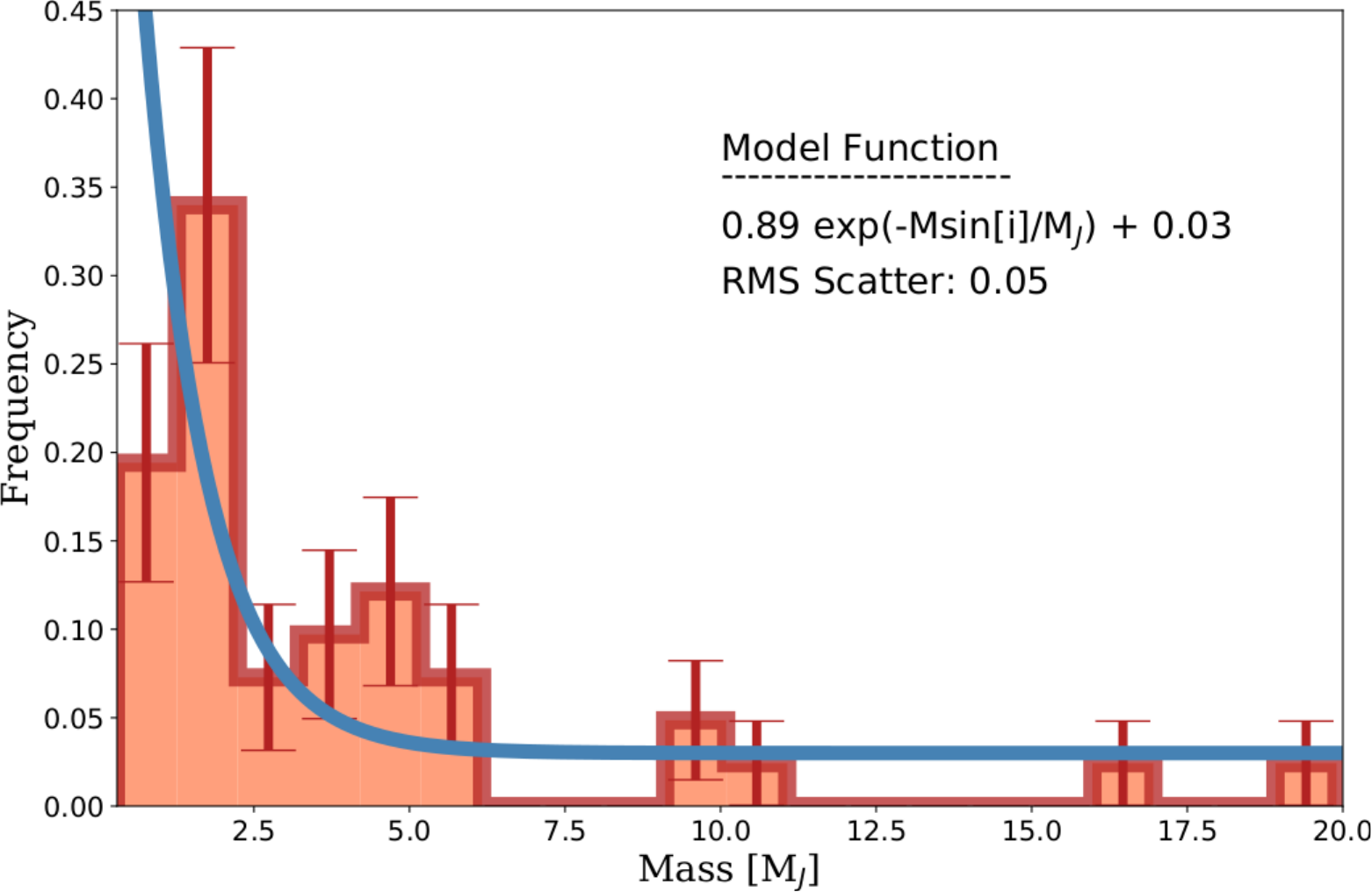}
\caption{Histogram showing the normalised frequency of giant planets orbiting low-luminosity giant stars, including those discovered in this work.  The associated uncertainties are drawn from Poisson statistics based on the individual counts within each bin.  The navy curve represents the exponential mass function proposed in \citet{jenkins17} for the giant planet sample orbiting lower-mass stars on the main sequence, with the model parameters are RMS scatter around this model, excluding the heavily incomplete bin, shown in the plot.  The planet candidates HIP24275b \& d, HIP75092c, and HIP90988b have been included in this analysis.}
\label{fig:mass_hist}
\end{figure}

It is interesting to explore the mass distribution of giant planets orbiting low-luminosity giant stars, particularly given that we are now beginning to uncover much lower-mass planets.  In Fig.~\ref{fig:mass_hist} we show a histogram of the normalised distribution of minimum masses, finding a strikingly similar form than what has been found for dwarf stars (e.g. \citealp{jorissen01}; \citealp{butler06}; \citealp{lopez12}).  The population is dominated by low-mass gas giants, those with minimum masses close to that of Jupiter.  Indeed, there seems to be an abundance of Jupiter-mass planets, and then a significant drop in the frequency for planets above $\sim$2~\mj, with a flat distribution above this value. Only a few examples of the highest-mass planets, (the easiest worlds to detect with this method for a given orbital period), have currently been found.  

If we compare this distribution to dwarf stars on the main sequence, we notice that a similar finding was also pointed out by \citet{jenkins17}, who found an exponential function to be an excellent description of the data.  The exponential model has the form:

\begin{equation}
    f(m) = A \times exp\left( -m~sin(i) \right) + B
\end{equation}

where A and B are constants that were found through a Markov chain Monte Carlo analysis in that work, having values of 0.89$\pm$0.03 and 0.030$^{+0.004}_{-0.003}$, respectively, and are highlighted in the plot.  Rather than fit a new exponential model to our distribution, given the fairly low number statistics, we overplot the distribution found by \citeauthor{jenkins17} for the dwarf stars in Fig.~\ref{fig:mass_hist} to test how well it describes the giant star sample.  The functional form does a good job of describing the minimum mass distribution of these giant planets, especially considering the different normalisation factors for each sample.  The RMS scatter around this model is found to be 0.05, after excluding the first bin that represents the lowest mass planets and is heavily incomplete.  Even including this bin only increases the scatter slightly, with a value of 0.07 found.  Such a result argues in favour of a single formation mechanism at play for both dwarf and giant star planetary systems.  Given that the progenitors of these giants were more massive than the typical solar mass dwarf star studies, (mostly A and F stars that are expected to have more massive proto-planetary disk dust mass fractions; \citealp{pascucci16}), it is reasonable to assume that core accretion dominates planet formation across all stellar masses in the inner systems, with only the outcomes differing in absolute terms, such as more massive planets orbiting more massive stars for example.  

\citet{jenkins17} discussed the existence of a pile-up of gas giant planets orbiting Sun-like stars with minimum masses above around 2~\mj\, indicating a break in the formation mechanism.  \citet{santos17} went on to study this in a more statistical sense, finding that indeed there is observational evidence for two distinct planet populations, where super-Jupiters with masses greater than $\sim$4~\mj\, tend to orbit more massive and more metal-poor hosts than lower-mass gas giants.  In Fig.~\ref{fig:mass_hist} we also see an emerging pile-up, or over-abundance of giant planets with minimum masses above 4~\mj. 

In order to test if there is statistical evidence for a second population here, we ran a Hartigan-like Dip Test (\citealp{hartigan85}).  Assuming the exponential model is a good description of the unimodal observed distribution of planets, we compared the maximum deviation of the cumulative distribution function (CDF) of the observed data against the maximum deviation of the CDF from 1M random realisations drawn from this model.  The model CDFs were scaled to match the maximum of the observed empirical CDF to take care of observational bias.  This allowed us to test to what extent the non-unimodality could appear due to random statistical fluctuations.  We found a deviation greater than the observed distribution occurs in $\sim$7\% of cases, meaning there is a $\sim$93\% probability that there exists a second population of super-Jupiters orbiting these giant stars.  Although this is not high enough to claim strong statistical significance at this time, by pushing towards 2$\sigma$ it is highly suggestive. 

As mentioned, the main sequence progenitors of the stars in the EXPRESS and PPPS samples were more massive than the typical Sun-like stars that make-up most dwarf star samples, and therefore we may expect the population of super-Jupiters to be more pronounced orbiting these stars, particularly the more metal-poor subset.  However, of the 13 planets with masses of $\sim$4~\mj\, or more discovered by these projects, only half of them orbit stars with sub-solar metallicities and of those half, only two are statistically separated from being in agreement with solar.  This is indicative that stellar mass is the dominant factor in separating the possible two populations of giant planets orbiting nearby stars, with any metallicity effect being only secondary.  Since a star's mass is correlated with its proto-planetary disk dust mass by a positive power index (\citealp{Andrews2013,pascucci16}), this second population of giant planets could be the result of a separate formation/migration pathway, that of gravitational collapse of a heavily fragmented disk (\citealp[e.g.][]{boss98,stamatellos15}).  However, to place this on a more robust statistical footing a fully homogeneous Bayesian analysis of the RVs for all stars in these projects should be performed, along with a similar homogeneous analysis of the stellar metallicities to help remove all sources of bias.  If this result holds, we can partition the population of gas giants into two based on their mass and formation channel, with Jupiters being formed through core accretion across all stellar masses from A$\--$Ks, and super-Jupiters ($>$2$\sim$3~\mj) being formed through gravitational collapse of massive unstable proto-planetary disks.

\section{Discussion and Summary}\label{sec8}

We have used precision RV data from the EXPRESS and PPPS giant star planet search projects to uncover two new nearby planetary systems, using a full Bayesian statistical approach on these types of stars.  HIP18606 and HIP111909 have been found to host three gas giant planets between them, with a single planet found orbiting HIP18606 on a 675~d orbit (HIP18606b) and having a minimum mass of $\sim$ 
0.8 \mj.  
HIP111909 on the other hand hosts two giant planets, both on long period orbits, with HIP111909b having an orbital period of nearly 500~d and HIP111909c orbiting close to 900~d.  Both minimum masses of these planets are in statistical agreement with that of Jupiter, adding to the large pantheon of Jupiter analogues orbiting stars that have recently evolved off of the main sequence, the so called low luminosity giant stars.  The period ratio of the planets is strikingly close to the 5:3 resonance, and some short dynamical tests hint at the possibility that they are currently in resonance, although more detailed dynamical modelling in the future is required to confirm such a scenario. 

In order to test the reliability and usefulness of combined Bayesian+MCMC approaches in the search for planets around giant stars, we also analysed an additional 11 low-luminosity giants, in order to confirm the reality of these detections, and to search for more planets in the systems.  Our approach confirmed all of the currently known planets orbiting these stars, which means that our modelling confirmed the signals are present. In addition, our activity index analysis, including new spectroscopic data, indicated that two signals could possibly be arising from stellar activity. In the case of HIP24275, the S-indices show a peak in the periodogram at a period close to that of planet c. However, the peak is found at low significance, and based on only 11 datapoints. Similarly, a periodogram of the NaD indicators of HIP90988 also reveals a peak close to the period of planet b, but again at low significance. We also found four new signals in the data, with the lowest amplitude of these possibly also originating from stellar activity, yet this takes the total number of planet candidates in this work to 20.  When we analyse these planets in the orbital period vs minimum mass plane, we find that the low-mass cohort are building the first vanguard of a new population of super-Saturns that orbit these stars beyond 200~d orbital periods.  Indeed, HIP\,75092c would be the lowest mass planet yet discovered by RV analysis orbiting any giant star.

Our Bayesian approach allowed a detailed picture to be brought forth on the underlying characteristics of the planetary orbits.  By statistically assessing models drawn from different eccentricity priors, we found that planets orbiting giant stars generally have circular orbits, indicating that the evolution of stars from the main sequence does not drastically affect widely separated giant planets.  This would render them excellent samples to directly compare with planets on the main sequence orbiting lower-mass stars.  The alternative is that their eccentricities are pumped throughout the stellar evolution off the main sequence, yet they quickly dissipate their orbital energy to return to circular orbits, possibly by interactions with other planets in the system.  In such a scenario, it becomes more complicated to directly compare them with the main sequence sample, at least in a quantitative way.  However, there is no global evidence as yet that this is indeed the case.

We could also test the mass function for these detected planets, and although the current numbers are somewhat small, an exponential function with values drawn from giant planet studies orbiting dwarf stars can explain well the distribution, albeit with a different normalisation factor.  These results lend weight to the reality of the core accretion model of planet formation being at play also in more massive stars than the Sun.  This analysis also revealed an emerging pile-up of super-Jupiters with minimum masses of $\sim 4\--5$\mj, a result in agreement with previous works performed on more Sun-like dwarf stars.  These results lend weight to the notion that there exists two separate populations of gas giant planets, and stellar mass, likely as a proxy for proto-planetary disk mass, is the dominant characteristic that drives this population bifurcation.

Our large grid of models told us that eccentricity prior width, not just the overall model, can be important when searching for true model parameters.  Set the prior too tight and the model does not have enough flexibility to search the wide parameter space efficiently, even if circular orbits are preferred.  Set the width too wide, and the parameter estimates can be questionable, along with the potential loss of real signals.  We also tested the usefulness of correlated noise models in these data, finding in general that white noise models do a better job of describing these types of RV data sets.  Regardless of the eccentricity prior used, models with only white noise or with a BIS linear correlation applied were statistically favoured over correlated noise models that included moving averages.  Therefore, random intrinsic variability of giant stars dominates over any correlated noise, at least at high frequency and with typical RV program sampling.  

Finally, additional emerging candidate signals were detected in these data sets, particularly for HIP111909, motivating the continued RV monitoring of these stars.  The statistical significance of these candidates never crossed our detection threshold, yet they were approaching relevance and were consistent across all models and model sets we tested.  In the coming years with more RVs, new candidates will likely be detected that further enhances the reputation of low-luminosity giant stars as rich hunting grounds for massive worlds, but we must caution that long-term stellar activity modelling is also required to identify which of these are real or not.  Even if we take the two previously announced planet candidates (HIP24275c and HIP90988b) that may be associated with activity index signals as being genuine false positives, then we could conclude that 12.5\% (2/16) of Doppler candidates orbiting these types of stars are not real, dropping the fraction of giant planets orbiting low-luminosity giant stars by the same factor, and hence moving down to around the 25$\--$30\% level.  
In \citet{vigan21}, the results presented from the direct imaging SHINE survey show that the fraction of widely separated (5-300~AU) substellar companions increases as a function of stellar mass, arriving at a value of 23.0$^{+13.5}_{-9.7}$\% for BA-type stars.  This is in good agreement with the value we have found here for at least a part of the parameter space that overlaps with our study.  Since the stars studied in \citeauthor{vigan21} are significantly younger than our sample and they reach out to much larger orbital separations, a hard comparison is beyond the scope of this work, but it is noteworthy that the results appear more than broadly consistent.

\section{Data Availability}

Appendices A, C, D, and E have been uploaded to the Zenodo website under the DOI 10.5281/zenodo.17024692, and can be found at the following link: https://zenodo.org/records/17024692.

\begin{acknowledgements}

We acknowledge the traditional owners of the land on which the AAT stands, the Gamilaraay people, and pay our respects to elders past and present.  We would like to thank the anonymous referee for their very thorough and helpful comments.  JSJ greatfully acknowledges support by FONDECYT grant 1240738 and from the ANID BASAL project FB210003.
R.B. acknowledges support from FONDECYT Project 1241963 and from ANID -- Millennium  Science  Initiative -- ICN12\_009.
C.A.G. acknowledges support from the National Agency for Research and Development (ANID) FONDECYT Postdoctoral Fellowship 2018 Project 3180668.  MRD is supported by CONICYT-PFCHA/Doctorado Nacional- 21140646, Chile.  J.C. is supported by a grant from the SC Space Grant Consortium.
M. T. P. acknowledges the support of Fondecyt-ANID fellowship ASTRON-0037.

\end{acknowledgements}

\bibliographystyle{aa}
\bibliography{refs}

\begin{appendix} 

\section{Stellar Activity Index Periodograms\label{appendixa}}

\section{Model Posterior Probabilities\label{appendixb}}
\begin{table*}[ht!]
\caption{HIP8541 BIC values for the best fit models for each of the examined model scenarios.}
\center
\scalebox{0.82}{
\begin{tabular}{ccccccc}
\hline
&&&Tightly Constrained Eccentricity\\
e$_{TC}$ Model & WNO & BLCO & BSLCO & MAO & BLCMA & BSLCMA \\
 & 411.1 (k1) & 414.0 (k1) & 417.2 (k1)  & 423.4 (k1)  & 428.7 (k1), [k2: 424.8] & 428.8 (k1) \\
\hline
&&&Constrained Eccentricity\\
e$_C$ Model & WNO & BLCO & BSLCO & MAO & BLCMA & BSLCMA \\
 & 410.8 (k1), [k2: 406.6] &  414.7 (k2)  &  408.5 (k2) &  422.9 (k1), [k2: 418.8]  & 421.5 (k2) & 430.2 (k1) [k2: 425.7] \\
\hline
&&&Fixed Eccentricity\\
e$_F$ Model & WNO & BLCO & BSLCO & MAO & BLCMA & BSLCMA \\
 & {\bf 394.6 (k2)}  & 397.9 (k2)  & 401.5 (k2)  &  406.0 (k2) & 409.9 (k2)  & 413.4 (k2) \\
\hline
\end{tabular}
}\\
Values in the curved brackets represent the model that the maximum posterior probability was drawn from.\\
Values in the square brackets represent the best preferred model fit, yet the signal was not confirmed in other model sets.\\
WNO: Model that includes White Noise Only.\\
BLCO: BIS Linear Correlation terms.\\
BSLCO: Combined BIS and S-index Linear Correlation terms.
\label{tab:8541bfs}
\end{table*}

\begin{table*}[h!]
\caption{HIP24275 BIC values for the best fit models for each of the examined model scenarios.}
\center
\scalebox{0.82}{
\begin{tabular}{ccccccc}
\hline
&&&Tightly Constrained Eccentricity\\
e$_{TC}$ Model & WNO & BLCO & BSLCO & MAO & BLCMA & BSLCMA \\
 & 531.7 (k3) & 540.4 (k2), [k3: 536.7] & 537.6 (k3)  & 549.6 (k2), [k3: 545.0]  & 553.6 (k2), [k3: 549.2] & 558.4 (k2), [k3: 553.6] \\
\hline
&&&Constrained Eccentricity\\
e$_C$ Model & WNO & BLCO & BSLCO & MAO & BLCMA & BSLCMA \\
 & 536.8 (k3) &  540.3 (k3)  &  541.7 (k3) &  550.8 (k3)  & 551.7 (k3) & 556.4 (k3) \\
\hline
&&&Fixed Eccentricity\\
e$_F$ Model & WNO & BLCO & BSLCO & MAO & BLCMA & BSLCMA \\
 & {\bf 514.0 (k2)}  & 518.0 (k2)  & 521.4 (k2)  &  528.0 (k2) & 531.7 (k2)  & 534.3 (k2) \\
\hline
\end{tabular}
}\\
Similar to the format of Table~\ref{tab:8541bfs} but for the star HIP\,24275.
\label{tab:8541bfs}
\end{table*}

\begin{table*}
\caption{HIP56640 BIC values for the best fit models for each of the examined model scenarios.}
\center
\begin{tabular}{ccccccc}
\hline
&&&Tightly Constrained Eccentricity\\
e$_{TC}$ Model & WNO & BLCO & BSLCO & MAO & BLCMA & BSLCMA \\
 & 220.7 (k1)  & 223.6 (k1) & 227.8 (k1) &  227.3 (k1) & 230.2 (k1) & 234.5 (k1) \\
\hline
&&&Constrained Eccentricity\\
e$_C$ Model & WNO & BLCO & BSLCO & MAO & BLCMA & BSLCMA \\
 & {\bf 214.9 (k1) } & 217.9 (k1) & 220.7 (k1) & 221.4 (k1) & 224.8 (k1) & 227.8 (k1) \\
\hline
&&&Fixed Eccentricity\\
e$_F$ Model & WNO & BLCO & BSLCO & MAO & BLCMA & BSLCMA \\
 & 218.6 (k1)  & 221.0 (k1) & 224.8 (k1) & 225.3 (k1)  & 227.5 (k1) & 231.7 (k1) \\
\hline
\end{tabular}\\
Similar to the format of Table~\ref{tab:8541bfs} but for the star HIP\,56640.

\label{tab:56640bfs}
\end{table*}

\begin{table*}
\caption{HIP67851 BIC values for the best fit models for each of the examined model scenarios\bf{, without the inclusion of Coralie data}.}
\center
\begin{tabular}{ccccccc}
\hline
&&&Tightly Constrained Eccentricity\\
e$_{TC}$ Model & WNO & BLCO & BSLCO & MAO & BLCMA & BSLCMA \\
 & 749.5 (k2) & 757.1 (k2) & 766.0 (k2)  &  764.7 (k2)  & 769.3 (k2) & 778.7 (k2) \\
\hline
&&&Constrained Eccentricity\\
e$_C$ Model & WNO & BLCO & BSLCO & MAO & BLCMA & BSLCMA \\
 & {\bf 746.3 (k2) } & 751.4 (k2) & 761.7 (k2)  &  759.9 (k2)  & 765.4 (k2) & 774.4 (k2) \\
\hline
&&&Fixed Eccentricity\\
e$_F$ Model & WNO & BLCO & BSLCO & MAO & BLCMA & BSLCMA \\
 & 751.9 (k2) & 756.6 (k2) & 763.4 (k2)  &  765.7 (k2)  & 771.0 (k2) & 777.8 (k2) \\
\hline
\end{tabular}\\
Similar to the format of Table~\ref{tab:8541bfs} but for the star HIP\,67851 without the inclusion of Coralie data.

\label{tab:67851bfs}
\end{table*}

\begin{table*}
\caption{HIP67851 BIC values for the best fit models for each of the examined model scenarios\bf{, with the inclusion of Coralie data}.}
\center
\begin{tabular}{ccccccc}
\hline
&&&Tightly Constrained Eccentricity\\
e$_{TC}$ Model & WNO & BLCO & BSLCO & MAO & BLCMA & BSLCMA \\
 & 1066.3 (k3) & 1078.3 (k3) & 1083.5 (k3)  &  1088.6 (k3)  & 1099.3 (k3) & 1098.6 (k3) \\
\hline
&&&Constrained Eccentricity\\
e$_C$ Model & WNO & BLCO & BSLCO & MAO & BLCMA & BSLCMA \\
 &  {\bf 1054.8 (k3)} & 1074.8 (k3) & 1081.4 (k3)   &  1075.3 (k3)  & 1087.6 (k3) & 1152.6 (k3) \\
\hline
&&&Fixed Eccentricity\\
e$_F$ Model & WNO & BLCO & BSLCO & MAO & BLCMA & BSLCMA \\
 & 1086.1 (k4) & 1093.8 (k4) & 1100.3 (k4)  &  1106.1 (k4)  & 1115.5 (k4) & 1123.3 (k4) \\
\hline
\end{tabular}\\
Similar to the format of Table~\ref{tab:8541bfs} but for the star HIP\,67851 with the inclusion of Coralie data.

\label{tab:67851bfs}
\end{table*}

\begin{table*}
\caption{HIP74890 BIC values for the best fit models for each of the examined model scenarios.}
\center
\begin{tabular}{ccccccc}
\hline
&&&Tightly Constrained Eccentricity\\
e$_{TC}$ Model & WNO & BLCO & BSLCO & MAO & BLCMA & BSLCMA \\
 & 263.3 (k1)  &  266.4 (k1) & 269.2 (k1) & 270.1 (k1) & 273.3 (k1) & 276.3 (k1) \\
\hline
&&&Constrained Eccentricity\\
e$_C$ Model & WNO & BLCO & BSLCO & MAO & BLCMA & BSLCMA \\
 & 262.3 (k1) & 265.9 (k1) & 268.0 (k1) & 269.6 (k1)  & 273.5 (k1) & 275.7 (k1) \\
\hline
&&&Fixed Eccentricity\\
e$_F$ Model & WNO & BLCO & BSLCO & MAO & BLCMA & BSLCMA \\
 & {\bf 258.3 (k1) } & 261.4 (k1) & 264.4 (k1) & 265.1 (k1)  & 268.2 (k1) & 271.6 (k1) \\
\hline
\end{tabular}\\
Similar to the format of Table~\ref{tab:8541bfs} but for the star HIP\,74890.

\label{tab:74890bfs}
\end{table*}

\begin{table*}
\caption{HIP75092 BIC values for the best fit models for each of the examined model scenarios.}
\center
\begin{tabular}{ccccccc}
\hline
&&&Tightly Constrained Eccentricity\\
e$_{TC}$ Model & WNO & BLCO & BSLCO & MAO & BLCMA & BSLCMA \\
 & 347.8 (k1)  &  349.5 (k1) & 354.9 (k1)  & 358.9 (k1) & 361.0 (k1) & 366.6 (k1) \\
\hline
&&&Constrained Eccentricity\\
e$_C$ Model & WNO & BLCO & BSLCO & MAO & BLCMA & BSLCMA \\
 & 347.7 (k1)  & 349.6 (k1) & 354.6 (k1) & 357.4 (k1) & 361.4 (k1) & 366.4 (k1) \\
\hline
&&&Fixed Eccentricity\\
e$_F$ Model & WNO & BLCO & BSLCO & MAO & BLCMA & BSLCMA \\
 & {\bf 334.0 (k2) } & 335.2 (k2) &  347.3 (k1) & 345.6 (k2) & 346.7 (k2) & 358.6 (k1) \\
\hline
\end{tabular}\\
Similar to the format of Table~\ref{tab:8541bfs} but for the star HIP\,75092.

\label{tab:75092bfs}
\end{table*}

\begin{table*}
\caption{HIP84056 BIC values for the best fit models for each of the examined model scenarios.}
\center
\begin{tabular}{ccccccc}
\hline
&&&Tightly Constrained Eccentricity\\
e$_{TC}$ Model & WNO & BLCO & BSLCO & MAO & BLCMA & BSLCMA \\
 & 556.2 (k1)  & 559.3 (k1)  & 561.6 (k1)  & 568.9 (k1)  & 572.0 (k1) & 574.0 (k1) \\
\hline
&&&Constrained Eccentricity\\
e$_C$ Model & WNO & BLCO & BSLCO & MAO & BLCMA & BSLCMA \\
 & 555.9 (k1) & 559.2 (k1) & 561.1 (k1) & 568.5 (k1) & 571.9 (k1) & 573.6 (k1) \\
\hline
&&&Fixed Eccentricity\\
e$_F$ Model & WNO & BLCO & BSLCO & MAO & BLCMA & BSLCMA \\
 & {\bf 550.5 (k1) }  & 553.3 (k1) & 556.7 (k1) &  563.1 (k1)  & 566.3 (k1) & 569.5 (k1) \\
\hline
\end{tabular}\\
Similar to the format of Table~\ref{tab:8541bfs} but for the star HIP\,84056.

\label{tab:84056bfs}
\end{table*}

\begin{table*}
\caption{HIP90988 BIC values for the best fit models for each of the examined model scenarios.}
\center
\begin{tabular}{ccccccc}
\hline
&&&Tightly Constrained Eccentricity\\
e$_{TC}$ Model & WNO & BLCO & BSLCO & MAO & BLCMA & BSLCMA \\
 & 353.5 (k2)  & 356.2 (k2) & 356.2 (k2) &  362.6 (k2)  & 371.6 (k2) & 377.7 (k2) \\
\hline
&&&Constrained Eccentricity\\
e$_C$ Model & WNO & BLCO & BSLCO & MAO & BLCMA & BSLCMA \\
 & 339.6 (k2)  & 364.0 (k2) & 346.1 (k2)  &  356.3 (k2)  & 370.4 (k2)  & 370.4 (k2) \\
\hline
&&&Fixed Eccentricity\\
e$_F$ Model & WNO & BLCO & BSLCO & MAO & BLCMA & BSLCMA \\
 & {\bf 339.3 (k2)} & 343.7 (k2) & 354.1 (k2)  &  351.4 (k2) & 355.8 (k2) & 355.8 (k2) \\
\hline
\end{tabular}\\
Similar to the format of Table~\ref{tab:8541bfs} but for the star HIP\,90988.

\label{tab:90988bfs}
\end{table*}

\begin{table*}
\caption{HIP95124 BIC values for the best fit models for each of the examined model scenarios.}
\center
\begin{tabular}{ccccccc}
\hline
&&&Tightly Constrained Eccentricity\\
e$_{TC}$ Model & WNO & BLCO & BSLCO & MAO & BLCMA & BSLCMA \\
 & 318.1 (k1) & 320.7 (k1) & 314.0 (k1)  & 329.6 (k1)  & 332.6 (k1) & 325.6 (k1) \\
\hline
&&&Constrained Eccentricity\\
e$_C$ Model & WNO & BLCO & BSLCO & MAO & BLCMA & BSLCMA \\
 & 311.5 (k1) & 320.7 (k1) & 315.6 (k1) & 330.9 (k1) & 333.6 (k1) & 326.5 (k1) \\
\hline
&&&Fixed Eccentricity\\
e$_F$ Model & WNO & BLCO & BSLCO & MAO & BLCMA & BSLCMA \\
 & 312.1 (k1) & 313.6 (k1) & {\bf 309.6 (k1)}  &  322.9 (k1) & 324.1 (k1) & 321.0 (k1) \\
\hline
\end{tabular}\\
Similar to the format of Table~\ref{tab:8541bfs} but for the star HIP\,95124.

\label{tab:95124bfs}
\end{table*}

\begin{table*}
\caption{HIP114933 BIC values for the best fit models for each of the examined model scenarios.}
\center
\begin{tabular}{ccccccc}
\hline
&&&Tightly Constrained Eccentricity\\
e$_{TC}$ Model & WNO & BLCO & BSLCO & MAO & BLCMA & BSLCMA \\
 & 337.2 (k1)  & 341.1 (k1) & 343.7 (k1) & 368.6 (k2)  & 373.1 (k2) & 355.4 (k1) \\
\hline
&&&Constrained Eccentricity\\
e$_C$ Model & WNO & BLCO & BSLCO & MAO & BLCMA & BSLCMA \\
 & 334.9 (k1)  &  338.2 (k1) & 341.0 (k1) & 346.4 (k1)  & 350.2 (k1) & 353.3 (k1) \\
\hline
&&&Fixed Eccentricity\\
e$_F$ Model & WNO & BLCO & BSLCO & MAO & BLCMA & BSLCMA \\
 & {\bf 333.6 (k1) } & 337.0 (k1) & 339.4 (k1) & 345.4 (k1) & 348.4 (k1) & 350.8 (k1) \\
\hline
\end{tabular}\\
Similar to the format of Table~\ref{tab:8541bfs} but for the star HIP\,114933.

\label{tab:114933bfs}
\end{table*}

\begin{table*}
\caption{HIP116630 BIC values for the best fit models for each of the examined model scenarios.}
\center
\begin{tabular}{ccccccc}
\hline
&&&Tightly Constrained Eccentricity\\
e$_{TC}$ Model & WNO & BLCO & BSLCO & MAO & BLCMA & BSLCMA \\
 & 298.9 (k1) & 299.6 (k1) & 306.1 (k1) & 309.8 (k1)  & 311.9 (k1) & 317.2 (k1) \\
\hline
&&&Constrained Eccentricity\\
e$_C$ Model & WNO & BLCO & BSLCO & MAO & BLCMA & BSLCMA \\
 & 297.7 (k1) & 298.9 (k1) & 305.2 (k1) &  308.7 (k1) & 310.7 (k1) & 316.5 (k1) \\
\hline
&&&Fixed Eccentricity\\
e$_F$ Model & WNO & BLCO & BSLCO & MAO & BLCMA & BSLCMA \\
 & {\bf 293.1 (k1) } & 294.1 (k1) & 300.2 (k1) & 304.6 (k1)  & 311.9 (k1) & 317.2 (k1) \\
\hline
\end{tabular}\\
Similar to the format of Table~\ref{tab:8541bfs} but for the star HIP\,116630.

\label{tab:116630bfs}
\end{table*}
\clearpage
\newpage

\section{Radial-velocity Analyses\label{appendixc}}

\section{ASAS Photometric Analyses\label{appendixd}}

\section{Hipparcos Photometric Analyses\label{appendixe}}

\section{Tables\label{appendixf}}

\small
\begin{table*}
\centering
\caption{Stellar properties}
\begin{tabular}{llllllll}
\hline\hline
\vspace{-0.3cm} \\
                  &   HIP\,8541   & HIP\,18606   & HIP\,24275    &  HIP\,56640   & HIP\,67851    &  HIP\,74890         & HIP\,75092   \\      
\hline \vspace{-0.3cm} \\
Spectral Type           & K2III/IV                  & K0/1III                    & K0III                     & K1III                     & K0III                      & K1III                 & K0III                      \\ 
$B$ (mag)               & 8.98$\pm$0.02             & 6.85$\pm$0.02              & 8.33$\pm$0.02             & 9.02$\pm$0.02             & 7.21$\pm$0.02              & 8.10$\pm$0.02         & 8.15$\pm$0.02              \\
$V$ (mag)               & 7.88$\pm$0.01             & 5.83$\pm$0.01              & 7.29$\pm$0.01             & 7.93$\pm$0.01             & 6.19$\pm$0.01              & 7.03$\pm$0.01          & 7.11$\pm$0.01              \\
$G$ (mag)               & 7.573$\pm$0.003           & 5.543$\pm$0.001            & 7.026$\pm$0.003           & 7.662$\pm$0.003         & 5.901$\pm$0.003          & 6.767$\pm$0.003      & 6.846$\pm$0.003        \\     
$J$ (mag)               & 6.01$\pm$0.02             & 4.25$\pm$0.26              & 5.57$\pm$0.02             & 6.13$\pm$0.02             & 4.48$\pm$0.32              & 5.29$\pm$0.02          & 5.36$\pm$0.02            \\ 
$H$ (mag)               & 5.46$\pm$0.04             & 3.62$\pm$0.22              & 5.13$\pm$0.04             & 5.64$\pm$0.02             & 3.81$\pm$0.25              & 4.81$\pm$0.02          & 4.36$\pm$0.02            \\
$K$ (mag)               & 5.29$\pm$0.03             & 3.60$\pm$0.26              & 4.95$\pm$0.02             & 5.49$\pm$0.02             & 3.75$\pm$0.24              & 4.64$\pm$0.02          & 4.74$\pm$0.03            \\
$\Pi$ (mas)             & 6.419$\pm$0.026           & 21.235$\pm$0.090           & 9.497$\pm$0.019           & 8.146$\pm$0.039           & 15.567$\pm$0.042           & 12.685$\pm$0.041       & 12.216$\pm$0.061           \\
\teff (K)               & 4672$\pm$61               & 4860$\pm$50                & 4909$\pm$61                 & 4769$\pm$55               & 4807$\pm$50                & 4824$\pm$71            & 4891$\pm$50                \\
log($L_\star$) (\lsun)  & 1.31$\pm$0.01             & 1.06$\pm$0.02              & 1.17$\pm$0.01             & 1.05$\pm$0.01             & 1.20$\pm$0.01              & 1.02$\pm$0.01          & 1.02$\pm$0.01   \\
\logg (\cmsq)           & 2.95$\pm$0.13             & 3.10$\pm$0.11              & 3.24$\pm$0.12             & 2.91$\pm$0.12             & 3.02$\pm$0.09              & 3.11$\pm$0.14          & 3.09$\pm$0.10              \\
\feh (dex)              & -0.19$\pm$0.06            & -0.01$\pm$0.05             & +0.13$\pm$0.06             & -0.03$\pm$0.05            & -0.07$\pm$0.05             & +0.12$\pm$0.06         & -0.01$\pm$0.05             \\
\vsini\, (k\ms)         & 2.90$\pm$0.92             & 3.04$\pm$0.51              & 3.44$\pm$0.56              & 3.09$\pm$0.68             & 3.33$\pm$0.49              & 3.21$\pm$0.74          & 3.13$\pm$0.59         \\
$M_\star$(\msun)        & 0.96$^{+0.06}_{-0.04}$    &  1.21$^{+0.11}_{-0.10}$    & 1.55$\pm$0.09              & 1.04$^{+0.07}_{-0.06}$    & 1.19$^{+0.10}_{-0.09}$     & 1.31$\pm$0.09          & 1.28$\pm$0.11     \\
\rstar(\rsun)           & 6.84$^{+0.08}_{-0.09}$    &  4.83$^{+0.13}_{-0.12}$    & 5.39$\pm$0.05              & 4.93$\pm$0.05             & 5.72$\pm$0.14              & 4.62$\pm$0.04          & 4.53$\pm$0.05     \\
age (Gyr)               & 10.97$^{+1.63}_{-1.97}$   & 5.29$^{+1.69}_{-1.31}$     & 2.42$^{+0.47}_{-0.37}$     & 9.15$^{+1.99}_{-1.89}$    & 5.46$^{+1.71}_{-1.31}$     & 4.41$^{+1.16}_{-0.90}$ & 4.30$\pm$1.26       \\
EEP                     & 508.7$^{+0.7}_{-0.8}$  & 495.0$^{+1.7}_{-1.1}$         & 497.1$^{+1.5}_{-1.9}$      & 497.5$^{+0.8}_{-0.7}$  & 500.8$^{+1.6}_{-1.1}$   & 493.1$\pm$0.5  &   492.6$\pm$0.6    \\
V$_{osc}$ (m/s)  &  21  &    10       &    10    &  11  &   13   &  8  &   8    \\

\vspace{-0.1cm} \\\hline\hline
\vspace{-0.3cm} \\

                 &   HIP\,84056                 & HIP\,90988                  &  HIP\,95124              & HIP\,114933    &  HIP\,116630         & HIP\,111909   \\
\hline \vspace{-0.3cm} \\
Spectral Type           &     K1III              &      K1III                 &    K0III                 &   K0III                    &   K1III               &   K1III                   \\  
$B$ (mag)               &  7.84$\pm$0.02           &  8.81$\pm$0.02             &  8.57$\pm$0.02           &  8.29$\pm$0.02             &  8.51$\pm$0.02       &   8.40$\pm$0.02             \\
$V$ (mag)               &  6.82$\pm$0.01            & 7.76$\pm$0.01              & 7.55$\pm$0.01             & 7.25$\pm$0.01              & 7.48$\pm$0.01         &  7.37$\pm$0.01            \\
$G$ (mag)               &  6.564$\pm$0.003           & 7.480$\pm$0.003           & 7.298$\pm$0.003        & 6.992$\pm$0.003            & 7.201$\pm$0.003       & 7.111$\pm$0.003        \\     
$J$ (mag)               &  5.07$\pm$0.04            &  6.00$\pm$0.02             & 5.85$\pm$0.02            &  5.56$\pm$0.02             &  5.75$\pm$0.02         & 5.65$\pm$0.02            \\ 
$H$ (mag)               &  4.73$\pm$0.08            &  5.53$\pm$0.03            &  5.37$\pm$0.03            &  5.05$\pm$0.02             & 5.30$\pm$0.05         & 5.27$\pm$0.03            \\
$K$ (mag)               &  4.49$\pm$0.04            &  5.38$\pm$0.02             & 5.25$\pm$0.02             & 4.91$\pm$0.02             & 5.13$\pm$0.03         & 5.02$\pm$0.02            \\
$\Pi$ (mas)             &  13.443$\pm$0.062       &  10.606$\pm$0.043     &  8.278$\pm$0.048       &   9.798$\pm$0.051      &  10.942$\pm$0.022     & 11.325$\pm$0.065           \\
\teff (K)               &   4880$\pm$50             &   4884$\pm$58   &    5008$\pm$50             &   4824$\pm$60              &    4879$\pm$61        & 4891$\pm$50                \\
log($L_\star$) (\lsun)  &   1.05$\pm$0.01        &  0.89$\pm$0.01          &    1.17$\pm$0.01         &     1.14$\pm$0.01          &  0.98$\pm$0.01       & 1.02$\pm$0.01   \\
\logg (\cmsq)           &  3.05$\pm$0.10         & 3.32$\pm$0.11              &   3.13$\pm$0.10           &   2.99$\pm$0.12            &   3.17$\pm$0.13        & 3.09$\pm$0.10              \\
\feh (dex)              &  -0.02$\pm$0.01          &  +0.17$\pm$0.06      &   +0.13$\pm$0.05          &    +0.06$\pm$0.06         &    +0.08$\pm$0.06     & -0.01$\pm$0.05             \\
\vsini\, (k\ms)         &  3.25$\pm$0.59       &    3.47$\pm$0.60           &  3.33$\pm$0.54           &  3.33$\pm$0.51          &  3.20$\pm$0.20      & 3.13$\pm$0.59         \\
$M_\star$(\msun)        &  1.21$\pm$0.09   &   1.30$\pm$0.08        &  1.64$\pm$0.09       &  1.39$\pm$0.09    &    1.28$\pm$0.09       & 1.28$\pm$0.11     \\
\rstar(\rsun)           &  4.72$\pm$0.07  &  3.94$\pm$0.04    &     5.31$\pm$0.06       &      5.27$\pm$0.05         &    4.28$\pm$0.05       & 4.53$\pm$0.05     \\
age (Gyr)               &  5.24$\pm$1.40      &  4.55$\pm$1.12        &   2.00$\pm$0.37    &   3.40$\pm$0.80   &   4.50$\pm$1.20     & 4.30$\pm$1.26       \\
EEP                     &  494.2$^{+1.3}_{-0.8}$ &    487.6$^{+0.6}_{-0.5}$   &  495.0$^{+1.9}_{-2.0}$ &  497.9$^{+0.6}_{-0.7}$  & 490.7$\pm$0.6  &   492.6$\pm$0.6    \\
V$_{osc}$ (m/s)  &  9  &    6       &    9    &  10  &   8   &  8       \\

\hline \vspace{-0.3cm} \\

\end{tabular}\\
EEP: Equivalent Evolutionary Point (see \citealp{dotter16} for details).\\
V$_{osc}$ is the expected RV pulsation velocity as explained in the text.
\label{tab:stellar_par}
\end{table*}
\normalsize

\begin{table}
\caption{Prior choices used in this work}
\centering
\begin{tabular}{ll}
\hline
Parameter                         & Prior                   \\ \hline
\\
Keplerian Parameters & \\ \hline
P [days]  & $\mathcal{U}(0.1, 3 \times max(t^1))$  \\
K [ms$^{-1}$]         & $\mathcal{U}(0.1, 3 \times max(|rv|))$ \\ 
$\Theta$ [rads]           & $\mathcal{U}(0, 2\pi)$       \\ 
$\omega$ [rads]         & $\mathcal{U}(0, 2\pi)$      \\ 
e$_{TC}$                 & $\mathcal{N}(0, 0.1^2)$            \\
e$_C$                 & $\mathcal{N}(0, 0.3^2)$            \\
e$_F$                 & $\mathcal{F}(0)$            \\  \hline
& \\
Noise Parameters                  &                                    \\ \hline
$\gamma$ [ms$^{-1}$]          & $\mathcal{U}(0, max(|rv|))$  \\  
$\theta_{jit}$ [ms$^{-1}$]    & $\mathcal{N}(5, 5)$       \\
Linear Correlation Coefficient $c$        & $\mathcal{U}(-max(|l|), max(|l|))$       \\ 
MA Coefficient $\phi$        & $\mathcal{U}(-1, 1)$       \\ 
MA Timescale $\tau$ [days]        & $\mathcal{F}(5)$       \\ \hline
& \\
Acceleration Parameter            &       \\     \hline                     
$\dot{\gamma}$ [ms$^{-1}$/yr]   & $\mathcal{U}(-1, 1)$    \\
\hline
\end{tabular}\\
$^1 t=$Time baseline of the data.\\
e$_{TC}$ and e$_{C}$ are the 'Tightly Constrained' and 'Constrained' priors, respectively.\\
$\mathcal{F}$ represents a fixed parameter.\\
Note that these are the general prior forms, used for the signal search.  All $k-1$ formats are constrained to the previously detected signal parameters (see text).
\label{tab:prior}
\end{table}

\begin{table}
\caption{Most probable \emp\, posterior mean RV model outcomes for HIP18606}
\center
\begin{tabular}{cc}
\hline
Parameter & e$_F$ WNO  \\ 
& k=1  \\\hline
P$_{orb}$ [d]     &   674.94$^{+6.43}_{-6.28}$    \\
Amplitude [m\,s$^{-1}$]     &  16.07$^{+1.12}_{-1.18}$  \\
Eccentricity      &  --   \\
$\omega$ [rads]    &  1.571   \\
$\Theta$ [rads]      &   4.47$^{+0.16}_{-0.14}$   \\
Msin($i$) [M$_{J}$]     &  0.77$\pm$0.07      \\
a [AU]    &   1.61$\pm$0.05    \\
$\dot{\gamma}$ [ms$^{-1}$yr$^{-1}$]  &  0.06$^{+0.56}_{-0.57}$   \\\hline

Nuisance Parameters &  \\
$\gamma_{UCLES}$ [m\,s$^{-1}$]  &   -3.54$\pm$2.08     \\
$\gamma_{Chiron_{pre}}$ [m\,s$^{-1}$]  &  -1.37$^{+3.25}_{-3.58}$  \\
$\gamma_{Chiron_{post}}$ [m\,s$^{-1}$]  &  1.59$^{+8.79}_{-8.58}$ \\
$\gamma_{FEROS}$ [m\,s$^{-1}$]  &  0.86$^{+4.66}_{-4.54}$   \\

$\theta_{jit,UCLES}$ [m\,s$^{-1}$]  &  3.93$^{+1.56}_{-1.13}$  \\
$\theta_{jit,Chiron_{pre}}$ [m\,s$^{-1}$]  &  6.42$^{+2.91}_{-2.40}$  \\
$\theta_{jit,Chiron_{post}}$ [m\,s$^{-1}$]  &  9.06$^{+3.86}_{-2.73}$  \\
$\theta_{jit,FEROS}$ [m\,s$^{-1}$]  &  9.59$^{+1.74}_{-1.50}$  \\

\hline
\end{tabular}\\
\small
The Chiron pre and post subscripts represent the data taken before and after the 2020 global pandemic SMARTS shutdown (October 26th, 2020).
\rm
\label{tab:18606}
\end{table}

\begin{table*}
\caption{HIP18606 posterior probability $\left(p\left(\vec{\theta} | D, M\right)\right)$ maxima for the best fit models for each of the examined model scenarios.}
\center
\begin{tabular}{ccccccc}
\hline
&&&Tightly Constrained Eccentricity\\
e$_{TC}$ Model & WNO & BLCO & BSLCO & MAO & BLCMA & BSLCMA \\
 & 428.0 (k1)  & 432.4 (k1) & 435.9 (k1)  & 443.8 (k1) & 449.0 (k1) & 451.6 (k1)  \\
\hline
&&&Constrained Eccentricity\\
e$_C$ Model & WNO & BLCO  & BSLCO & MAO & BLCMA & BSLCMA \\
 & 427.2 (k1) & 431.4 (k1)  & 435.9 (k1)  & 443.5 (k1)  & 447.7 (k1) & 452.3 (k1)  \\
\hline
&&&Fixed Eccentricity\\
e$_F$ Model & WNO & BLCO & BSLCO & MAO & BLCMA & BSLCMA \\
 & {\bf 419.5 (k1) }  & 424.0 (k1)  & 427.9 (k1) & 436.0 (k1) & 439.9 (k1) & 444.2 (k1)  \\
\hline
\end{tabular}\\
Values in the curved brackets represent the model that the maximum posterior probability was drawn from.\\
WNO: Model that includes White Noise Only.\\
BLCO: BIS Linear Correlation terms.\\
BSLCO: Combined BIS and S-index Linear Correlation terms.\\
\label{tab:18606bfs}
\end{table*}

\begin{table}
\caption{\texttt{EMPEROR} posterior best RV model outcome for HIP111909}
\center
\begin{tabular}{ccc}
\hline
Parameter &  e$_F$ WNO \\
& k=1  & k=2 \\\hline
P$_{orb}$ [d]     &  487.08$^{+3.81}_{-3.63}$ & 893.63$^{+14.89}_{-16.03}$ \\
Amplitude [m\,s$^{-1}$]     &  25.57$^{+2.03}_{-1.52}$ & 13.86$^{+1.25}_{-1.14}$\\
Eccentricity      &  -- & -- \\
$\omega$ [rads]    & 1.571  & 1.571 \\
$\Theta$ [rads]      &  4.70$^{+0.16}_{-0.19}$  & 0.59$^{+0.29}_{-0.25}$ \\
Msin($i$) [M$_{J}$]     &  1.21$\pm$0.10 & 0.81$\pm$0.08  \\
a [AU]    &   1.32$^{+0.04}_{-0.05}$ & 1.97$^{+0.08}_{-0.03}$ \\
$\dot{\gamma}$ [ms$^{-1}$yr$^{-1}$]  &  -0.76$^{+0.78}_{-0.71}$  &   \\\hline
Nuisance Parameters &&\\
$\gamma_{UCLES}$ [m\,s$^{-1}$]  &  -17.04$^{+2.74}_{-3.53}$  &  \\
$\gamma_{Chiron_{pre}}$ [m\,s$^{-1}$]  &   -7.38$^{+5.18}_{-5.54}$ &  \\
$\gamma_{Chiron_{post}}$ [m\,s$^{-1}$]  &   -8.40$^{+13.19}_{-9.96}$ & \\
$\gamma_{FEROS}$ [m\,s$^{-1}$]  &  5.59$^{+4.44}_{-4.34}$  &  \\

$\theta_{jit,UCLES}$ [m\,s$^{-1}$]  &  6.11$^{+1.70}_{-1.34}$  &   \\
$\theta_{jit,Chiron_{pre}}$ [m\,s$^{-1}$]  &  4.29$^{+2.17}_{-2.08}$  &   \\
$\theta_{jit,Chiron_{post}}$ [m\,s$^{-1}$]  &  14.26$^{+2.41}_{-2.52}$  &  \\
$\theta_{jit,FEROS}$ [m\,s$^{-1}$]  &  12.17$^{+1.93}_{-1.52}$  &  \\

\hline
\end{tabular}\\
\small
The Chiron pre and post subscripts represent the data taken before and after the 2020 global pandemic SMARTS shutdown (October 26th, 2020).
\rm
\label{tab:111909}
\end{table}

\begin{table*}
\caption{HIP111909 posterior probability $\left(p\left(\vec{\theta} | D, M\right)\right)$ maxima for the best fit models for each of the examined model scenarios.}
\center
\begin{tabular}{ccccccc}
\hline
&&&Tightly Constrained Eccentricity\\
e$_{TC}$ Model & WNO & BLCO & BSLCO & MAO & BLCMA & BSLCMA \\
 & 473.2 (k2)  & 473.1 (k2)  & 479.0 (k2)  & 489.4 (k2) & 493.3 (k2)  & 497.1 (k2) \\
\hline
&&&Constrained Eccentricity\\
e$_C$ Model & WNO & BLCO & BSLCO & MAO & BLCMA & BSLCMA \\
 & 469.2 (k2)  & 472.7 (k2) &  477.4 (k2)  & 485.3 (k2) & 490.1 (k2) & 490.8 (k2) \\
\hline
&&&Fixed Eccentricity\\
e$_F$ Model & WNO & BLCO & BSLCO & MAO & BLCMA & BSLCMA \\
 & {\bf 457.6 (k2)}  & 461.5 (k2)  & 465.6 (k2)  & 473.0 (k2) & 478.3 (k2) & 481.0 (k2) \\
\hline
\end{tabular}\\
Values in the curved brackets represent the model that the maximum posterior probability was drawn from.\\
Values in the square brackets represent the best preferred model fit, yet the signal was not confirmed in other model sets.\\
WNO: Model that includes White Noise Only.\\
BLCO: BIS Linear Correlation terms.\\
BSLCO: Combined BIS and S-index Linear Correlation terms.\\
\label{tab:111909bfs}
\end{table*}

\begin{sidewaystable*}
\tiny
\caption{\emp\, orbital and physical parameters for the stars with known planets studied in this work, showing the results from the statistically favoured model only.}
\center
\begin{tabular}{ccccccccccccc} 
\hline
Planet & P$_{orb}$  & Amplitude & Eccentricity & $\omega$ & $\Theta$ & Msin($i$) & $\dot\gamma$ &  $\theta_{jit,UCLES}$ & $\theta_{jit,Chiron}$ & $\theta_{jit,FEROS}$ & $\theta_{jit,Hires}$ & $\theta_{jit,Coralie}$  \\ 
 & [d]  & [m\,s$^{-1}$] & & [rads] & [rads] & [M$_{Jup}$] & [m\,s$^{-1}$\,yr$^{-1}$] & [m\,s$^{-1}$]  & [m\,s$^{-1}$] & [m\,s$^{-1}$] & [m\,s$^{-1}$]  \\ 
 \hline
HIP\,8541b$^1$ &  1583.72$^{+21.90}_{-18.43}$ & 85.01$^{+3.06}_{-4.16}$  & 0 & 0 & 2.06$^{+0.09}_{-0.07}$  & 4.71$^{+0.28}_{-0.27}$  & -1.67$^{+2.18}_{-2.21}$  & 3.41$^{+3.57}_{-2.68}$ & 9.95$^{+1.79}_{-1.52}$  & 9.04$^{+2.71}_{-2.13}$ & -- & -- \\  
HIP\,8541c$^1$ &  714.23$^{+13.85}_{-13.50}$  & 16.06$^{+1.44}_{-1.50}$ & 0 & 0 &  4.50$^{+0.27}_{-0.25}$ & 0.70$\pm$0.07  & -- & -- & -- & -- & -- & -- \\ 
HIP\,24275b$^1$ &  551.73$^{+3.68}_{-3.85}$ & 31.67$^{+0.80}_{-1.41}$  & 0 & 0 & 0.67$^{+0.12}_{-0.11}$  & 4.71$^{+1.66}_{-0.08}$  & 0.77$^{+0.74}_{-0.71}$  & 4.89$^{+1.34}_{-1.11}$ & 9.95$^{+1.79}_{-1.52}$  & 9.04$^{+2.71}_{-2.13}$ & 7.61$^{+1.13}_{-0.91}$ & -- \\  
HIP\,24275c$^1$ &  904.66$^{+15.15}_{-14.28}$  & 23.33$^{+1.47}_{-1.21}$ & 0 & 0 &  5.85$\pm$0.17 & 1.47$\pm$0.10  & -- & -- & -- & -- & -- & -- \\ 
HIP\,56640b$^2$ & 2999.33$^{+123.88}_{-99.11}$ &  54.66$^{+2.20}_{-2.08}$ & 0.166$^{+0.008}_{-0.007}$  & 1.744$^{+0.025}_{-0.018}$ &  1.664$^{+0.186}_{-0.175}$  & 3.92$\pm$0.26 &  2.94$^{+1.38}_{-1.26}$  & 0.96$^{+3.96}_{-2.97}$ & $\--$ &   7.59$^{+1.56}_{-1.37}$ & -- & --  \\ 
HIP\,67851b$_{NC}^4$ & 89.18$\pm$0.07  &  49.27$^{+0.88}_{-0.07}$  & 0.102$\pm$0.006 & 3.138$^{+0.003}_{-0.029}$  & 0.27$\pm$0.08  & 1.20$^{+0.08}_{-0.07}$  &  -5.17$^{+0.03}_{-0.36}$  & 14.46$^{+2.89}_{-1.47}$  & 6.78$^{+1.28}_{-1.05}$  & 5.90$^{+1.13}_{-1.32}$  & -- & -- \\  
HIP\,67851c$_{NC}^3$ & 2372.52$\pm$34.23  & 65.44$^{+1.92}_{-1.97}$  & 0.133$\pm$0.007  & 2.605$^{+0.027}_{-0.026}$  & 2.14$^{+0.06}_{-0.07}$  & 4.75$^{+0.30}_{-0.32}$  & -- & -- & -- & -- & -- & -- \\  
HIP\,67851b$_{C}^4$ & 89.11$^{+0.05}_{-0.07}$  &  45.97$^{+1.40}_{-0.88}$  & 0.0001$\pm$0.0002 & 3.140$^{+0.003}_{-0.24}$  & 5.18$\pm$0.18  & 1.14$\pm$0.07  &  1.84$^{+0.50}_{-0.54}$  & 9.69$^{+2.49}_{-1.24}$  & 6.34$^{+1.20}_{-1.01}$  & 7.35$^{+1.80}_{-1.13}$  & -- & 14.18$^{+2.00}_{-1.47}$ \\  
HIP\,67851c$_{C}^3$ & 3268.56$^{+51.89}_{-41.37}$  & 58.40$^{+1.00}_{-0.18}$  & 0.252$^{+0.008}_{-0.011}$  & 1.063$^{+0.022}_{-0.016}$  & 2.17$\pm$0.07  & 4.67$^{+0.26}_{-0.29}$  & -- & -- & -- & -- & -- & -- \\  
HIP\,67851d$_{C}^3$ & 1816.31$^{+15.92}_{-18.20}$  & 35.01$^{+1.37}_{-3.31}$  & 0.589$\pm$0.015  & 3.135$^{+0.006}_{-0.008}$  & 2.80$\pm$0.10  & 1.82$^{+0.18}_{-0.15}$  & -- & -- & -- & -- & -- & -- \\  
HIP\,74890b$^1$ & 812.47$^{+19.07}_{-11.84}$  & 35.50$^{+2.17}_{-2.07}$  & 0 & 0 & 5.85$^{+0.21}_{-0.22}$ & 1.92$^{+0.16}_{-0.14}$  &  -32.43$^{+1.25}_{-1.36}$ & 6.95$^{+2.22}_{-1.65}$  & -- & 9.84$^{+2.06}_{-1.65}$  & -- & -- \\  
HIP\,75092b$^2$ & 970.78$^{+14.71}_{-15.71}$ & 26.56$^{+1.81}_{-1.37}$ & 0 & 0 & 2.85$^{+0.16}_{-0.18}$ & 1.52$^{+0.13}_{-0.12}$  &  -1.60$^{+0.70}_{-0.80}$ & 5.06$^{+2.32}_{-2.53}$  & 2.52$^{+4.65}_{-3.54}$  & 10.04$^{+2.03}_{-1.51}$   & -- & -- \\  
HIP\,75092c$^2$ & 207.23$^{+1.28}_{-1.34}$ & 9.90$^{+0.88}_{-0.87}$ & 0 & 0 & 5.45$\pm$0.31  & 0.33$^{+0.04}_{-0.03}$ & -- & -- & -- & --  & -- & -- \\  
HIP\,84056b$^1$ & 842.81$^{+12.61}_{-11.46}$  & 35.87$^{+2.18}_{-2.13}$  & 0 & 0 & 2.65$^{+0.15}_{-0.14}$  & 1.88$^{+0.15}_{-0.14}$ & 6.61$^{+1.52}_{-1.41}$  &  13.81$^{+2.33}_{-1.94}$  & 11.02$^{+2.20}_{-1.44}$  & 7.57$^{+1.99}_{-1.63}$  & -- & -- \\  
HIP\,90988b$^2$ & 452.45$^{+2.17}_{-1.97}$ & 38.77$^{+2.04}_{-2.15}$  & 0 & 0 & 2.75$^{+0.12}_{-0.10}$ & 1.71$^{+0.12}_{-0.11}$  &  -52.04$^{+0.76}_{-0.44}$ & 13.99$^{+3.34}_{-3.20}$  & 3.24$^{+2.54}_{-2.31}$  & 5.56$^{+1.58}_{-1.26}$  & -- & -- \\  
HIP\,90988c$^2$ & 2561.12$^{+131.97}_{-69.46}$ & 18.96$\pm$1.50  & 0 & 0 & 1.84$^{+0.19}_{-0.20}$  & 1.41$^{+0.13}_{-0.14}$  & -- & -- & -- & -- & -- & -- \\  
HIP\,95124b$^1$ & 563.60$^{+5.71}_{-3.72}$ & 42.61$^{+2.08}_{-2.30}$  & 0 & 0 & 1.71$\pm$0.14 & 2.40$^{+0.15}_{-0.16}$  &  -7.49$^{+2.28}_{-2.28}$  & 13.53$^{+4.01}_{-2.89}$  & 4.00$^{+2.35}_{-2.01}$  &  8.53$^{+2.19}_{-1.30}$  & -- & -- \\  
HIP\,114933b$^2$ & 1479.03$^{+43.43}_{-44.42}$  & 24.90$^{+1.43}_{-1.50}$  & 0 & 0 & 2.46$^{+0.20}_{-0.18}$  & 1.76$\pm$0.13  & -0.69$^{+1.06}_{-1.11}$  & 5.77$^{+2.54}_{-1.67}$  & 3.44$^{+2.09}_{-2.18}$  & 10.57$^{+2.03}_{-1.68}$  & -- & -- \\  
HIP\,116630b$^5$ & 873.04$^{+14.70}_{-15.50}$ & 30.83$^{+1.75}_{-1.84}$  & 0 & 0 & 3.76$^{+0.16}_{-0.12}$ & 1.70$\pm$0.13  & 0.04$^{+1.07}_{-1.14}$  & 4.61$^{+2.17}_{-1.54}$  & 4.56$^{+3.48}_{-2.99}$  & 6.38$^{+1.69}_{-1.47}$  & -- & -- \\  
\hline
\end{tabular}\\
Original planetary detections: 1) \citealp{jones16}; 2) \citealp{jones21}; 3) \citealp{jones15a}; 4) \citealp{jones15b}; 5) \citealp{wittenmyer17}\\
Subscripts C and NC refer to the analyses conducted with (C=Coralie) and without (NC=No Coralie) the inclusion of Coralie RV data, respectively\\

\label{tab:confirm}
\end{sidewaystable*}

\end{appendix}
\end{document}